\documentclass[preprint2]{aastex}

\slugcomment{Not to appear in Nonlearned J., 45.}

\begin{document}


\title{The Effects of Thermonuclear Reaction-Rate Variations on $^{26}$Al Production in Massive Stars: a Sensitivity Study}


\author{Christian Iliadis\altaffilmark{1,2}, Art Champagne\altaffilmark{1,2}, Alessandro Chieffi\altaffilmark{3} and Marco Limongi\altaffilmark{4}}
\altaffiltext{1}{Department of Physics and Astronomy, University of North Carolina at Chapel Hill, Chapel Hill, NC 27599-3255, USA; iliadis@unc.edu, aec@tunl.duke.edu.}
\altaffiltext{2}{Triangle Universities Nuclear Laboratory, Durham, NC 27708-0308, USA.}
\altaffiltext{3}{Istituto Nazionale di Astrofisica - Istituto di Astrofisica Spaziale e Fisica Cosmica, Via Fosso del Cavaliere, I-00133, Roma, Italy; alessandro.chieffi@iasf-roma.inaf.it.}
\altaffiltext{4}{Istituto Nazionale di Astrofisica - Osservatorio Astronomico di Roma, Via Frascati 33, I-00040, Monteporzio Catone, Italy; marco@oa-roma.inaf.it.}



\begin{abstract}
We investigate the effects of thermonuclear reaction rate variations on $^{26}$Al production in massive stars. The dominant production sites in such events were recently investigated by using stellar model calculations: explosive neon-carbon burning, convective shell carbon burning, and convective core hydrogen burning. Post-processing nucleosynthesis calculations are performed for each of these sites by adopting temperature-density-time profiles from recent stellar evolution models. For each profile, we individually multiplied the rates of all relevant reactions by factors of 10, 2, 0.5 and 0.1, and analyzed the resulting abundance changes of $^{26}$Al. In total, we performed $\approx 900$ nuclear reaction network calculations. Our simulations are based on a next-generation nuclear physics library, called STARLIB, which contains a recent evaluation of Monte Carlo reaction rates. Particular attention is paid to quantifying the rate uncertainties of those reactions that most sensitively influence $^{26}$Al production.  For stellar modelers our results indicate to what degree predictions of $^{26}$Al nucleosynthesis depend on currently uncertain nuclear physics input, while for nuclear experimentalists our results represent a guide for future measurements. 
We also investigate equilibration effects of $^{26}$Al. In all previous massive star investigations, either a single species or two species of $^{26}$Al were taken into account, depending on whether thermal equilibrium was achieved or not. These are two extreme assumptions and in a hot stellar plasma the ground and isomeric state may communicate via $\gamma$-ray transitions involving higher-lying $^{26}$Al levels.

We tabulate the results of our reaction rate sensitivity study for each of the three distinct massive star sites referred to above. It is found that several current reaction rate uncertainties influence the production of $^{26}$Al. Particularly important reactions are $^{26}$Al(n,p)$^{26}$Mg, $^{25}$Mg($\alpha$,n)$^{28}$Si, $^{24}$Mg(n,$\gamma$)$^{25}$Mg and $^{23}$Na($\alpha$,p)$^{26}$Mg. These reactions should be prime targets for future measurements. Overall, we estimate that the nuclear physics uncertainty of the $^{26}$Al yield predicted by the massive star models explored here amounts to about a factor of 3. We also find that taking the equilibration of $^{26}$Al levels explicitly into account in any of the massive star sites investigated here has only minor effects on the predicted $^{26}$Al yields. Furthermore, we provide for the interested reader detailed comments regarding the current status of certain reactions, including $^{12}$C($^{12}$C,n)$^{23}$Mg, $^{23}$Na($\alpha$,p)$^{26}$Mg, $^{25}$Mg($\alpha$,n)$^{28}$Si, $^{26}$Al$^{m}$(p,$\gamma$)$^{27}$Si, $^{26}$Al(n,p)$^{26}$Mg and $^{26}$Al(n,$\alpha$)$^{23}$Na. 
\end{abstract}





\section{Introduction\label{intro}}
The radioisotope $^{26}$Al is of outstanding importance for $\gamma$-ray astronomy and cosmochemistry. 
It has been discovered in three distinct sites: (i) in the Galactic interstellar medium via detection of its decay emission line at 1809 keV (Mahoney et al. 1982, Diehl et al. 1995); (ii) in meteorites via observed excesses of its radioactive decay (daughter) product $^{26}$Mg (MacPherson, Davies \& Zinner 1995), implying an injection of live $^{26}$Al into the early Solar System nebula; and (iii) in presolar dust grains, again via detected $^{26}$Mg excesses (Hoppe et al. 1994; Huss, Hutcheon \& Wasserburg 1997), that are uncontaminated by solar system material and thus are of likely stellar origin. Identification of the main sources of $^{26}$Al would have far-reaching implications, ranging from questions related to the circumstances and conditions of the Solar System birth to imposing strong constraints on the chemical evolution of the Galaxy. A number of different sources have been suggested over the years: AGB stars, classical novae, Wolf-Rayet stars, and core collapse supernovae. For reviews, see Prantzos \& Diehl (1996) or Diehl \& Timmes (1998). However, the origin of $^{26}$Al remains controversial. 

The observation of Galactic $\gamma$-rays from $^{26}$Al is important since it provides unambiguous {\it direct} evidence for the theory of nucleosynthesis in stars. The half-life of $^{26}$Al amounts to $7.17\times10^5$ y and is small compared to the time scale of Galactic chemical evolution ($\approx$10$^{10}$ y). Consequently, nucleosynthesis is currently occurring in the interstellar medium and, in particular, $^{26}$Al is synthesized throughout the Galaxy. From the observed $\gamma$-ray intensity, depending on the assumption for the density distribution, a present-day $^{26}$Al equilibrium mass of $\approx 2-3$~$M_\odot$ in the entire Galaxy has been inferred (Diehl et al. 2006). The observational evidence favors in this case massive stars as a source: first, the all-sky map of the 1809 keV $\gamma$-ray line detected by the COMPTEL instrument onboard CGRO showed that $^{26}$Al is confined along the Galactic disk and that the measured intensity is clumpy and asymmetric (Pl\"uschke et al. 2001); second, the comparison between the $^{26}$Al all-sky map from COMPTEL to other all-sky maps for different wavelengths (Kn\"odlseder 1999) revealed that the 1.8 MeV $\gamma$-ray emission is correlated with the Galactic free-free emission, which traces the distribution of ionized gas in the interstellar medium, observed in the microwave domain by the COBE satellite; third, the measurement of the 1809 keV line Doppler shift by the SPI spectrometer onboard INTEGRAL demonstrated that $^{26}$Al co-rotates with the Galaxy and hence supports a Galaxy-wide origin (Diehl et al. 2006). Here we investigate the bulk production of $^{26}$Al in massive stars. A study of $^{26}$Al/$^{27}$Al ratios observed in meteorites and presolar grains will be subject of a separate study.

Massive stars may produce $^{26}$Al during several different phases of their evolution: (i) during pre-supernova stages in the C/Ne convective shell, where a fraction of the $^{26}$Al survives the subsequent explosion and is ejected into the interstellar medium (Arnett \& Wefel 1978); (ii) during core collapse via explosive Ne/C burning (Arnett 1977), where the ejected $^{26}$Al yield may perhaps be modified by the $\nu$-process via neutrino spallation (Woosley et al. 1990); and (iii) in Wolf-Rayet stars, i.e., stars with masses in excess of about $30~M_{\odot}$, which experience such a strong mass loss that even layers located within the H convective core, hence significantly enriched in $^{26}$Al, are ejected into the interstellar medium (Palacios et al. 2005). These $^{26}$Al production mechanisms (with the exception of the $\nu$-process) were recently analyzed in detail by Limongi \& Chieffi (2006) by using extensive stellar model calculations of solar metallicity stars in the mass range of $11M_\odot \leq M \leq 120M_\odot$. In that work they also emphasized the impact of rate uncertainties for selected reactions on the final $^{26}$Al yields. Another discussion along these lines can be found in Woosley \& Heger (2007).

In the present work we expand this effort by presenting a comprehensive investigation of the impact of nuclear reaction rate uncertainties on the synthesis of $^{26}$Al. Our method is similar in spirit to earlier nuclear reaction rate sensitivity studies that addressed the nucleosynthesis in classical novae (Iliadis et al. 2002) and type I x-ray bursts (Parikh et al. 2008). The general strategy consists of varying the rates of many reactions by different factors (in this work, 10, 2, 0.5 and 0.1) and to analyze the impact of each individual reaction rate change on the final $^{26}$Al yields. Once the yield changes are established for this grid of rate variation factors, more realistic abundance changes based on actual rate uncertainties, if available, are considered. At present it is not feasible to perform this computationally intensive procedure with a self-consistent stellar model. Instead, we extract representative temperature-density-time profiles from recent stellar evolution models of massive stars and execute a large number of post-processing reaction network sensitivity calculations using these profiles. Our goal is two-fold. On the one hand, we would like to quantify to what degree predictions of $^{26}$Al yields depend on currently uncertain nuclear physics input. On the other hand, by identifying the ``most important nuclear reactions", our results represent a guide for future measurements. 

There are a number of novel aspects about the present work. First, we employ a new-generation library of nuclear reaction and weak interaction rates, called STARLIB. This library is partially based on a recent evaluation of experimental Monte Carlo reaction rates (Iliadis et al. 2010). Besides recommended reaction rates for a grid of temperature values between 1 MK and 10 GK, the library includes in addition for many reactions the rate uncertainty factor at each temperature. In fact, this work represents the first application of STARLIB. Second, we carefully investigate the equilibration effects of $^{26}$Al. At least two species of $^{26}$Al take part in the nucleosynthesis, the ground state and the isomeric state. In all previous investigations, either a single species or two species of $^{26}$Al were taken into account, depending on whether thermal equilibrium is achieved or not. Obviously, these are two extreme assumptions and in a hot stellar plasma the ground and isomeric state may ``communicate" via $\gamma$-ray transitions involving higher-lying $^{26}$Al levels.

Our study has some obvious limitations. First, since we perform post-processing calculations, we necessarily focus our investigation on the effects of nuclear reaction rates. In other words, the important effects of convection\footnote{It can be shown analytically that the abundance evolution in a convective region in which the turnover time is fast enough to ensure a flat abundance profile of the various nuclear species is equivalent to the evolution of a single mesh in which each local thermonuclear rate, $\left< \sigma v \right>_{ij}$, is replaced by its mass-weighted average over the convective region, i.e., $\sum_k \left< \sigma v \right>_{ij,k}~dm_k/m_{tot}$. Hence it is perfectly plausible to use a single point evolution for a convective region if the turnover time is faster that the nuclear burning time (e.g., for core H burning). Furthermore, we performed the calculation using un-weighted rates instead of mass-weighted average rates because we are mostly interested in $^{26}$Al abundance {\it changes}.}, mass loss, rotation, and so on, are outside the scope of the present work. This also implies that our simulations are unsuitable for defining {\it absolute} $^{26}$Al yields. Instead, we claim that our procedure is useful for exploring the effects of $^{26}$Al abundance {\it changes} that result from reaction rate variations. Second, we only explore a few temperature-density-time evolutions that are representative of solar metallicity stars. A more comprehensive study covering a broad range of stellar masses and metallicities is also beyond the scope of this work. Third, it is well-known that the radioisotope $^{60}$Fe (half-life of $2.62\times 10^6$ y) is likely co-produced with $^{26}$Al in massive stars  and that their abundance ratio provides a sensitive constraint on stellar models (see Limongi \& Chieffi 2006; Woosley \& Heger 2007; and references therein). Indeed, Galactic $\gamma$-rays from the decay of $^{60}$Fe have been detected by both RHESSI and the SPI spectrometer onboard INTEGRAL, and the observed $\gamma$-ray line flux ratio for $^{60}$Fe and $^{26}$Al amounts to $\approx 0.1-0.2$ (Harris et al. 2005). In the present work we only focus on the nucleosynthesis of $^{26}$Al and leave a similar sensitivity study for $^{60}$Fe to future work\footnote{In the recent work of Tur, Heger \& Austin (2010), the impact of triple-$\alpha$ and $^{12}$C($\alpha$,$\gamma$)$^{16}$O rate variations on the $^{26}$Al, $^{44}$Ti and $^{60}$Fe yields were investigated using stellar evolution and explosion models. The authors claimed that ``...over a range of twice the experimental uncertainty, $\sigma$, for each helium-burning rate, the production of $^{26}$Al, $^{60}$Fe, and their ratio vary by factors of 5 or more...". By looking in detail at their results, it is clear that the effects on the $^{60}$Fe yield are indeed large. However, it becomes also apparent that the effects on the $^{26}$Al yield are much smaller. For example, their 25$M_\odot$ model and adopting the initial abundances from Lodders (2003) provides a factor of 1.5 change in $^{26}$Al yield if the triple-$\alpha$ and $^{12}$C($\alpha$,$\gamma$)$^{16}$O rates are individually varied by their 1$\sigma$ experimental uncertainties (see their Tab. 3). As will be seen, the rate uncertainty effects explored in the present work result in significantly larger $^{26}$Al yield variations.}. Finally, by individually varying each rate and leaving all other rates at their nominal values, we disregard any correlations among different reactions. In our opinion, no single study will cover all of the possible uncertainties, but each approach has advantages and disadvantages. We present only one realization of a sensitivity study, similar to the procedure applied in Iliadis et al. (2002) and Parikh et al. (2008). Interestingly, the latter work explored two methods: the one applied here and also a Monte Carlo procedure. It was found by Parikh et al. (2008) that very similar results were obtained with ``...minor differences attributed to such correlation effects...". We feel that a Monte Carlo procedure makes most sense if it is performed with {\it reliable} experimentally based reaction rate probability densities. However, as will become apparent below, we do not have this information for all of the important reaction rates yet and thus leave such a study to future work when an update of STARLIB becomes available.

This paper is organized as follows. Our procedure is explained in more detail in $\S$ 2, including a discussion of stellar models, the equilibration of $^{26}$Al, and a description of the library STARLIB. The results of reaction rate sensitivity studies are presented in $\S$ 3 for the three predicted main sites of $^{26}$Al synthesis: explosive Ne/C burning, convective C/Ne shell burning, and convective H core burning. A summary and conclusions are given in $\S$ 4. More information on reaction and decay rates, together with a discussion of individual reactions, is provided in the Appendix.

\section{General Procedure}
\subsection{Massive star models\label{starmodel}}
The stellar models adopted in the present work are those presented in Limongi \& Chieffi (2006). For the sake of completeness, we summarize the basic properties of these models and the main results concerning the production of $^{26}$Al in massive stars. The evolution of each stellar model was computed from the pre-main sequence phase up to the onset of the iron core collapse by using the stellar evolutionary code FRANEC (Frascati RAphson Newton Evolutionary Code, release 5.050218). The kernel of this code has been presented in Limongi \& Chieffi (2003) (and references therein). Here we will only mention recent updates. First, the convective mixing and the nuclear burning were coupled together providing a set of diffusion equations that are linearized and solved simultaneously by means of a Newton-Raphson method. This coupling is extremely important in all situations where the nuclear burning timescale of a given nuclide is comparable to the mixing turnover time. Thus the interaction between the local nuclear burning and the convective mixing cannot be disregarded. The nuclear network adopted was the same as in Limongi \& Chieffi (2003) and the thermonuclear reaction rates were up-to-date at the time when these models were computed (see Tab. 1 of Limongi \& Chieffi 2006). The nuclide $^{26}$Al was treated in a distinct manner, by assuming two separate species (for the ground and isomeric state) for temperatures below $T\approx 1$~GK, and a single (thermalized) species above this temperature. Mass loss was included following the prescriptions of Vink et al. (2000) for the blue supergiant phase ($T_{eff}>12000$~K), de Jager et al. (1988) for the red supergiant phase ($T_{eff}<12000$~K), and Nugis \& Lamers (2000) for the Wolf-Rayet phase. All of these solar metallicity models had an initial He mass fraction of 0.285 and a global metallicity (by mass) of $Z=0.02$. The relative abundances for the various nuclear species were adopted from Anders \& Grevesse (1989).

The explosion of the mantle of the star was started artificially, by instantaneously imparting an initial velocity of $v_{0}$ to a mass coordinate corresponding to $\approx 1~M_\odot$ of the presupernova model. Such a mass coordinate relates to a region located well within the iron core and is chosen in such a way that the initial conditions should not affect the properties of the shock wave too much at a time when it approaches the Fe-Si interface. The formation and propagation of the shock wave, generated in such a way, was calculated by means of a computer code that solves the fully compressible reactive hydrodynamic equations by applying the Piecewise Parabolic Method (PPM) of Colella \& Woodward (1984), using a Lagrangian scheme. The chemical evolution of the matter was computed by coupling the same nuclear reaction network that was adopted in the hydrostatic calculations to the system of hydrodynamic equations. The free parameter $v_{0}$ was properly adjusted in order to obtain a given final kinetic energy of the ejecta or, equivalently, to eject a given amount of mass above the Fe core. Since $^{26}$Al was synthesized in regions relatively far away from the Fe core (see below), its final yield did not depend on the particular choice of $v_{0}$ provided that at least a minimum amount of $^{56}$Ni was ejected.

Based on this set of presupernova models and simulated explosions, it was found that $^{26}$Al is produced in massive stars in three distinct evolutionary stages: core H burning, C convective shell burning just prior to the core collapse, and explosive Ne/C burning. Any $^{26}$Al produced by these massive stars is ejected into the interstellar medium, both by stellar winds and the explosion, in different proportions depending on the initial stellar mass. It was also found that $^{26}$Al was mainly produced by explosive Ne/C burning over most of the initial mass interval between 11 and 120~ $M_{\odot}$. Only for the more massive stars, say, $M>60 M_{\odot}$, did the wind component (produced during core H burning) become important (see Tab. 3 and Fig. 2 in Limongi \& Chieffi 2006 for details). In the present work, we based our study on 20, 60 and 80 $M_\odot$ model stars for exploring $^{26}$Al yield sensitivities in explosive Ne/C burning, C convective shell burning and core H burning, respectively. These choices of model stars are motivated by the fact that they provide relatively large $^{26}$Al yields.
 
\subsection{Nuclear physics library\label{physlib}}
The nuclear physics input for the present post-processing studies is based on a new-generation library, called STARLIB. It originated from a previous version of REACLIB (originally created by F.-K. Thielemann) that one of us modified over the years and was used for all of the reaction network calculations presented in Iliadis 2007. 

At that point in time several important changes occurred. A recent evaluation of reaction rates for the A=14-40 target range (Iliadis et al. 2010) was completed. These 62 {\it experimental} rates are based on a Monte Carlo technique, allowing for a rigorous definition of recommended reaction rates and their associated uncertainties. The Monte Carlo procedure also provides, for the first time, for any given temperature the (output) reaction rate probability density function that is based on the (input) probability densities of measured nuclear physics quantities (such as S-factors, resonance energies, resonance strengths, upper limits in spectroscopic factors, etc.). From the cumulative distributions of the rate probability densities, a low rate, median rate and high rate can be defined as the 0.16, 0.50 and 0.84 quantiles, respectively, assuming a coverage probability of 68\%. The meaning of these rates is in general different from the commonly reported, but statistically meaningless, literature expressions ``lower limit", ``nominal value" and ``upper limit" of the total reaction rate. It is important to emphasize that the Monte Carlo rates incorporate both statistical {\it and} systematic uncertainties, as explained in detail in Longland et al. (2010). Furthermore, it has been shown in Longland et al. (2010) that in the majority of cases the Monte Carlo rate probability density function can be approximated by a lognormal distribution, which is determined by only two parameters: the lognormal location parameter $\mu$ and the lognormal spread parameter $\sigma$. The former parameter determines the recommended reaction rate via $N_A\left< \sigma v \right>_{rec} = e^{\mu}$, while the latter parameter corresponds to the rate factor uncertainty via $f.u. = e^{\sigma}$. 

The information on the rate probability density was not available previously and opens interesting windows of opportunity for Monte Carlo studies of nucleosynthesis and energy generation in stars. However, it becomes clear from the above discussion that three quantities ($T$, $N_A\left< \sigma v \right>_{rec}$, lognormal $\sigma$) instead of the traditional two ($T$ and $N_A\left< \sigma v \right>_{rec}$) need to be reported so that the user can calculate the rate probability density for each reaction at each temperature. Therefore, it was decided to convert our 2007 version of the REACLIB, which lists the recommended reaction rates as analytical functions of temperature by employing a number of rate fitting parameters, to a tabular format. To be precise, the rate tables are directly derived from the fitting parameters and not from any tabular rates given in the original publications. The new format consists of three columns and lists for each reaction the temperature, the recommended rate and the rate factor uncertainty on a grid of 60 temperatures between 1 MK and 10 GK, allowing for an accurate interpolation between grid points. At this stage the rate factor uncertainty for each reaction is set equal to a nominal value of 10. In a subsequent step, the rates and factor uncertainties of 62 reactions in the A=14-40 region were replaced with their exact Monte Carlo results. In addition, the rates of the following interactions were replaced with more recent information\footnote{Assuming that the reaction rate probability density function can be approximated by a lognormal distribution, it can be shown (Iliadis et al. 2010) that for a coverage probability of 68\% the lognormal spread parameter is given by $f.u. = e^\sigma = \sqrt{x_{high}/x_{low}}$, where $x_i$ denotes a reaction rate; this expression can be employed to derive approximate rate factor uncertainties, $f.u.$, from published high and low reaction rate boundaries. Furthermore, all of the replaced rates are obtained from the exact rate tables of the original publications, not from any fitting parameters.}: (i) 10 Big Bang reactions, using the rates of Descouvemont et al. (2004), which were derived from an R-matrix description of the available data; (ii) 30 reactions from the NACRE evaluation of experimental rates (Angulo et al. 1999), in the mass range of A=1-26; (iii) (n,$\gamma$) reactions based on the KADoNiS v0.2 evaluation of experimental rates (Dillmann et al. 2006); (iv) a number of special reactions, such as $^{14}$N(p,$\gamma$)$^{15}$O (Bertone et al., in preparation) and $^{12}$C($\alpha$,$\gamma$)$^{16}$O (Kunz et al. 2002); (v) 550 experimental rates for $\beta$-decays and $\beta$-delayed particle decays including associated uncertainties, calculated from the half lives and branching ratios compiled in Audi et al. (2003); and (vi) 17 $\gamma$-ray transitions rates for $^{26}$Al (see below). For all nuclear reactions mentioned above, the corresponding reverse reaction rates were also calculated and properly accounted for in the library. Furthermore, all experimental reaction rates were corrected for the effects of thermal target excitations using the stellar enhancement factors and partition functions of Rauscher \& Thielemann (2000), although it should be noted that these effects are relatively small for the sites of nucleosynthesis discussed here. For all other reactions for which insufficient experimental information is available to compute reliable experimental rates, the results of statistical model calculations (Rauscher \& Thielemann 2000) were adopted. 

The new library, STARLIB, described above extends in its present version up to antimony (Sb) and is employed for the very first time in the present work. A more detailed account of STARLIB will be published elsewhere (Iliadis et al., in preparation). We emphasize that the rate probability density functions contained in STARLIB are not directly used in the present post-processing calculations. However, the tabulated rate factor uncertainties will be useful in later sections for the discussion of reaction rate uncertainties.

\subsection{Thermal equilibration of $^{26}$Al\label{thermequi}}
\begin{figure}
\includegraphics[scale=.50]{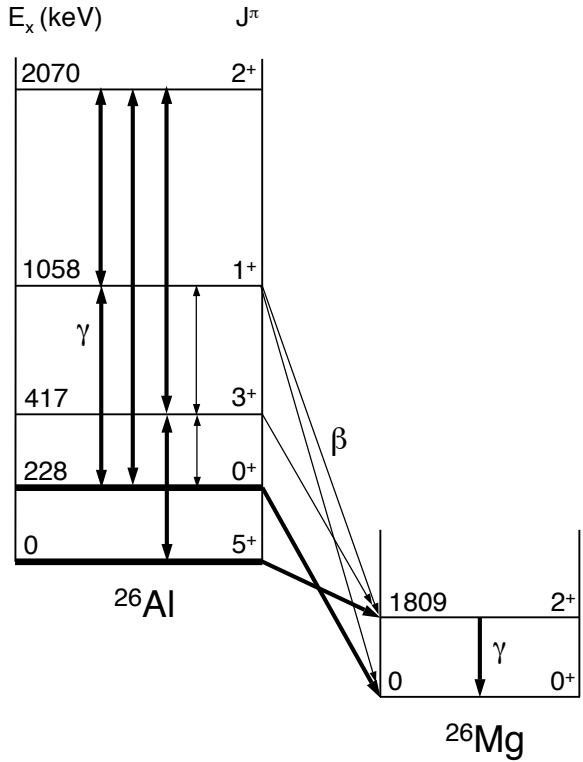}
\caption{Level scheme of $^{26}$Al. Energies and $J^\pi$-values are from Endt (1990). The vertical arrows represent $\gamma$-ray transitions. In $^{26}$Al, the thick vertical lines denote experimentally measured transitions (Endt 1990), while the decay rates for the thin vertical lines have been estimated using shell model calculations (Coc, Porquet \& Nowacki 1999; Runkle, Champagne \& Engel 2001). The possible communication of the $^{26}$Al ground state and isomeric state (228 keV) via thermal excitations involving higher-lying levels (417, 1058 and 2070 keV) is apparent. The arrows connecting $^{26}$Al and $^{26}$Mg denote $\beta$-decay transitions, according to the shell model calculations of Kajino et al. (1988). The $\beta$-decays shown do not necessarily represent direct transitions, but rather indicate if the ground state or first excited state (at 1809 keV) in $^{26}$Mg is predominantly populated by a given transition.\label{figlevel}}
\end{figure}

A level scheme of $^{26}$Al is shown in Fig. \ref{figlevel}. The 5$^+$ ground state, $^{26}$Al$^g$, $\beta$-decays with a half-life of T$_{1/2}=7.17\times10^5$ y to several excited states (not shown in the figure) in the $^{26}$Mg daughter nucleus that mainly de-excite via $\gamma$-ray transitions to the first excited state (at 1809 keV) in $^{26}$Mg. The subsequent decay to the $^{26}$Mg ground state gives rise to the 1809 keV $\gamma$-ray line emission observed in the Galactic plane ($\S$ \ref{intro}). An interesting situation occurs in $^{26}$Al since its first excited state (228 keV; 0$^+$), $^{26}$Al$^m$, is an isomer. In other words, the significant angular momentum difference between ground and isomeric state gives rise to a very small $\gamma$-ray decay constant (with a mean lifetime on the order of $\approx 10^6$ y for a M5 decay). Instead, the isomer $\beta$-decays to the ground state of $^{26}$Mg (without emission of a $\gamma$-ray) with a half-life of T$_{1/2}=6.34$ s.

Although the direct $\gamma$-ray transition between $^{26}$Al$^g$ and $^{26}$Al$^m$ is strongly inhibited, they may nevertheless be linked (or ``communicate") with each other via thermal excitations involving higher-lying $^{26}$Al levels. Three of these levels are shown in Fig. \ref{figlevel}, at excitation energies of 417 keV ($^{26}$Al$^a$), 1058 keV ($^{26}$Al$^b$) and 2070 keV ($^{26}$Al$^c$). For example, the ground and isomeric state are linked via the $\gamma$-ray transitions $0\leftrightarrow417\leftrightarrow228$. It is obvious that at sufficiently high temperatures and long timescales $^{26}$Al$^g$ and $^{26}$Al$^m$ will achieve thermal equilibrium, implying that their abundance ratio is simply determined by the Boltzmann distribution (i.e., by plasma temperature and energy difference), and that the internal equilibration mechanism is not important. However, it is also clear that, by lowering temperature and time scale, the ground and isomeric state will fall out of thermal equilibrium at some point. The important issue to consider for a reaction network calculation is if and when $^{26}$Al should be treated as a single species ($^{26}$Al$^t$, denoting thermal equilibrium) or as two distinct species ($^{26}$Al$^g$ and $^{26}$Al$^m$) that $\beta$-decay with their characteristic laboratory half lives. This issue was investigated by Ward \& Fowler (1980), who established the generally accepted procedure: above a limiting temperature value of T=0.4 GK, the nuclide $^{26}$Al should be regarded as a single species (with all of its levels in thermal equilibrium), while below this temperature two distinct species (no equilibrium) should be assumed. 

Some of the important $\gamma$-ray transition rates in $^{26}$Al have not been measured yet (indicated by the thin vertical arrows in Fig. \ref{figlevel}) and Ward \& Fowler (1980) had to employ for these rather crude approximations. In more recent work (Coc, Porquet \& Nowacki 1999; Runkle, Champagne \& Engel 2001) the $\gamma$-ray transition rates have been estimated by using shell model calculations and the thermal equilibration of $^{26}$Al was studied in more detail using small reaction networks. These works confirmed the assumption that above $T\approx0.45$ GK the ground and isomeric state are in thermal equilibrium, while below $T\approx0.15$ GK these two states $\beta$-decay with their laboratory half lives (i.e., neither equilibrium nor any communication between levels). In the transitional temperature region, $T\approx0.15-0.45$ GK, the effective $^{26}$Al decay rate was found to deviate substantially from the results reported by Ward \& Fowler (1980) by up to 4 orders-of-magnitude and, furthermore, that $^{26}$Al$^g$ and $^{26}$Al$^m$ are indirectly linked by $\gamma$-ray transitions (i.e., they achieve a quasi-equilibrium). It was pointed out by Runkle, Champagne \& Engel (2001) that these findings rest on simplifying assumptions and may not hold if the nuclear reactions that produce or destroy $^{26}$Al are sufficiently fast to disturb the thermal equilibration. See also, Gupta \& Meyer (2001).

In the present work we examine carefully the equilibration of $^{26}$Al for each of the three nucleosynthesis sites mentioned above. Two separate post-processing calculations using recommended interaction rates are performed and the resulting $^{26}$Al yields are compared: one assuming either a single or two separate $^{26}$Al species, depending on the temperature regime (according to Ward \& Fowler 1980), and one where the communication between ground and isomeric states is explicitly taken into account. For the latter case, no artificial assumptions about the equilibration of $^{26}$Al are made, but additional $^{26}$Al species (i.e., levels at 417, 1058 and 2070 keV; see Fig. \ref{figlevel}) need to be taken into account in the reaction network. The $\gamma$-ray transition rates linking the various levels in $^{26}$Al, and the most important $\beta$-decay transition rates to $^{26}$Mg, are adopted from Runkle, Champagne \& Engel (2001). Since these have not been published elsewhere, we list the decay constants versus temperature in Appendix \ref{appdec}. Furthermore, when taking more than two $^{26}$Al species into account, the reaction rates for $^{26}$Al$^{g}$ and $^{26}$Al$^m$ are separately needed. This represents an additional complication since, as will be seen below, some of these rates are poorly known at present. When performing the two separate post-processing calculations referred to above, it is of utmost importance that an {\it internally consistent} set of rates for the production and destruction of $^{26}$Al$^t$, $^{26}$Al$^g$ and $^{26}$Al$^{m}$ is employed. Suppose, for example, that the rate of a destruction reaction $^{26}$Al$^x$(a,b) in the first calculation (i.e., assuming a single $^{26}$Al species) is adopted from one specific source, and the rates of the physically related reactions $^{26}$Al$^g$(a,b) and $^{26}$Al$^m$(a,b) in the second calculation (i.e., assuming five different $^{26}$Al species) are adopted from a different source. In such a case, when comparing the results of the two network calculations, one may find significant differences in $^{26}$Al yields. However, these may not be caused at all by the effects of thermal equilibration of $^{26}$Al but may rather reflect spurious results caused by using inconsistent reaction rates. In order to elucidate this issue, we list in Appendix \ref{appreac} the rates and sources of all reactions considered in our network that produce and destroy $^{26}$Al.

\section{Procedure and Results}
\subsection{General considerations\label{gencons}}
The same reaction network is used in the present work for investigating the nucleosynthesis of $^{26}$Al in the predicted main locations of massive stars: explosive Ne/C burning, convective shell C/Ne burning and convective core H burning. The network extends from $^{1}$H to $^{40}$Ca, including 175 proton- and neutron-rich nuclides up to the respective driplines, that are linked by 1648 nuclear interactions ($\S$ \ref{physlib}). Initial abundances are listed in Tab. \ref{tababun} and the temperature-density time evolutions for each of the sites will be discussed in the following subsections. The nucleosynthesis will initially be visualized by considering so-called ``abundance flows", which represent the change of abundance per time as a result of an interaction between two nuclides. Since a forward and corresponding reverse reaction occur concurrently, what is of main interest is the {\it net} abundance flow (i.e., the difference between forward and reverse flow). Initially, a ``standard" network calculation is performed, employing recommended reaction rates. The final abundances, achieved at the end of the standard calculations, are summarized in Tab. \ref{tababun}. Subsequently, a series of network calculations is performed, where the rates of many reactions are varied individually by generic factors of 10, 2, 0.5 and 0.1. Resulting abundance changes of $^{26}$Al are then analyzed in detail. Then we focus our discussion on the actual rate uncertainties in the most relevant temperature region, which differs from site to site. It will become apparent that this temperature region is rather narrow, which significantly reduces any (unknown) systematic effects that are potentially caused by an incorrect temperature dependence of some rates. Finally, the issue of thermal equilibration of $^{26}$Al is investigated.
\begin{deluxetable}{ccccccccc}
\tabletypesize{\scriptsize}
\tablewidth{0pt}
\tablecaption{Initial and final mass fractions of present post-processing calculations\tablenotemark{a}\label{tababun}} 
\tablehead{ \colhead{Nuclide} & \multicolumn{3}{c}{$X_i$\tablenotemark{b}} & & \multicolumn{3}{c}{$X_f$\tablenotemark{c}} & Solar\tablenotemark{d} \nl
\cline{2-4}
\cline{6-8}
\colhead{} & \colhead{xNe/C} & \colhead{C/Ne} & \colhead{H} & & \colhead{xNe/C} & \colhead{C/Ne} &
             \colhead{H} & \colhead{}
}
\startdata
$^{1}$H    & \nodata  & \nodata & 7.0E-01 && \nodata  & \nodata & 1.4E-06 & 7.11E-01 \nl
$^{2}$H    & \nodata  & \nodata & 5.0E-05 && \nodata  & \nodata & \nodata & 2.76E-05 \nl
$^{3}$He  & \nodata  & \nodata & 3.1E-05 && \nodata  & \nodata & \nodata & 3.40E-05 \nl
$^{4}$He  & \nodata  & \nodata & 2.9E-01 && \nodata  & \nodata & 9.8E-01 & 2.74E-01 \nl
$^{12}$C  & 1.9E-02 & 1.5E-01 & 3.2E-03 && 1.5E-02  & 1.0E-01 & 2.6E-04 & 2.44E-03 \nl
$^{13}$C  & \nodata & \nodata & 3.8E-05 && \nodata  & \nodata & 7.3E-05 & 2.96E-05 \nl
$^{14}$C  & \nodata & 1.5E-05 & \nodata && \nodata   & 2.7E-06  & \nodata & \nodata \nl
$^{14}$N  & 8.2E-07 & 1.2E-04 & 1.2E-03 && \nodata   & 2.0E-05 & 1.3E-02& 7.90E-04 \nl
$^{15}$N  & \nodata & \nodata & 4.6E-06 && \nodata   & \nodata & \nodata & 3.11E-06 \nl
$^{16}$O  & 5.0E-01 & 7.0E-01 & 1.0E-02 && 5.8E-01  & 6.7E-01 & 1.0E-04 & 6.55E-03 \nl
$^{17}$O  & \nodata & 2.9E-05 & 4.1E-06 && \nodata   & 3.7E-06 & 4.0E-07 & 2.60E-06 \nl
$^{18}$O  & \nodata & \nodata & 2.3E-05 && \nodata   & \nodata & \nodata & 1.48E-05 \nl
$^{20}$Ne & 3.4E-01 & 1.0E-01 & 1.7E-03 && 1.8E-01  & 1.7E-01 & 1.6E-03 & 1.16E-03 \nl
$^{21}$Ne & 2.6E-04 & 7.0E-04 & 4.3E-06 && 3.2E-06  & 1.2E-04 & \nodata & 2.92E-06 \nl
$^{22}$Ne & 1.4E-04 & 1.0E-03 & 1.4E-04 && \nodata  & 1.0E-04 & 2.9E-05 & 9.37E-05 \nl
$^{22}$Na & 2.1E-06 & 4.2E-06 & \nodata && 1.4E-07  & 4.4E-07 & \nodata & \nodata  \nl
$^{23}$Na & 7.5E-03 & 9.6E-03 & 3.5E-05 && 3.5E-04  & 8.5E-03 & 2.7E-04 & 3.84E-05 \nl
$^{24}$Mg & 4.6E-02 & 4.6E-03 & 5.4E-04 && 8.7E-02 & 1.2E-02 & 5.4E-04 & 5.61E-04 \nl
$^{25}$Mg & 1.0E-02 & 5.9E-03 & 7.1E-05 && 2.2E-03 & 6.1E-03 & 3.8E-06 & 7.40E-05 \nl
$^{26}$Mg & 8.7E-03 & 9.4E-03 & 8.1E-05 && 2.1E-03 & 1.1E-02 &6.0E-05 & 8.47E-05 \nl
$^{26}$Al  & 3.5E-06 & 5.8E-07 & \nodata && 1.3E-04  & 2.4E-05 & 4.9E-05 & \nodata \nl
$^{27}$Al  & 1.1E-02 & 8.6E-03 & 6.1E-05 && 1.1E-02  & 9.5E-03 & 1.1E-04 & 6.59E-05 \nl
$^{28}$Si  & 3.7E-02 & 1.1E-03 & 6.9E-04 && 6.7E-02  & 2.3E-03 & 6.9E-04 & 7.49E-04 \nl
$^{29}$Si  & 1.4E-03 & 2.3E-04 & 3.6E-05 && 1.2E-02  & 6.8E-04 & 3.6E-05 & 3.94E-05 \nl
$^{30}$Si  & 6.7E-04 & 1.8E-04 & 2.5E-05 && 1.2E-02  & 3.6E-04 & 2.5E-05 & 2.69E-05 \nl
$^{31}$P  & 3.9E-04 & 8.3E-05 & 8.6E-06 && 2.5E-03  & 1.0E-04 & 8.6E-06 & 7.53E-06 \nl
$^{32}$P  & 1.5E-06 & 1.6E-06 & \nodata && 3.4E-06  & 3.2E-06 & \nodata & \nodata \nl
$^{33}$P  & 3.3E-07 & 7.3E-07 & \nodata && 2.1E-06  & 1.4E-06 & \nodata & \nodata \nl
$^{32}$S  & 1.4E-02 & 2.4E-04 & 4.2E-04 && 1.6E-02  & 2.2E-04 & 4.2E-04 & 3.92E-04 \nl
$^{33}$S  & 1.9E-04 & 2.7E-05 & 3.4E-06 && 3.8E-04  & 5.2E-06 & 3.4E-06 & 3.20E-06 \nl
$^{34}$S  & 2.7E-03 & 1.0E-04 & 2.0E-05 && 1.9E-03  & 1.0E-04 & 2.0E-05 & 1.85E-05 \nl
$^{35}$S  & 1.3E-06 & 6.9E-07 & \nodata && 3.1E-05  & 2.0E-06 & \nodata & \nodata \nl
$^{36}$S  & 1.9E-06 & 9.5E-07 & \nodata && 3.5E-05  & 2.2E-06 & \nodata & 7.93E-08 \nl
$^{35}$Cl  & 2.9E-04 & 4.9E-06 & 2.7E-06 && 1.6E-03 & 7.1E-06 & 2.7E-06 & 4.03E-06 \nl
$^{36}$Cl  & 7.7E-07 & 3.1E-07 & \nodata && 2.1E-05 & 2.8E-07 & \nodata & \nodata \nl
$^{37}$Cl  & 3.4E-05 & 5.1E-05 & 9.0E-07 && 4.1E-05 & 4.9E-05 & 9.2E-07 & 1.36E-06 \nl
$^{36}$Ar  & 9.6E-04 & 1.7E-05 & 8.1E-05 && 8.8E-04 & 1.3E-05 & 8.1E-05 & 9.05E-05 \nl
$^{37}$Ar  & 1.2E-05 & \nodata & \nodata && 1.5E-05  & \nodata & \nodata & \nodata \nl
$^{38}$Ar  & 1.5E-03 & 2.7E-05 & 1.6E-05 && 8.2E-04 & 2.9E-05 & 1.6E-05 & 1.74E-05 \nl
$^{39}$Ar  & 2.9E-06 & 8.5E-07 & \nodata && 1.1E-05 & 2.6E-06 & \nodata & \nodata \nl
$^{39}$K  & 1.3E-04 & 4.7E-06 & 3.6E-06 && 7.3E-04  & 4.2E-06 & 3.6E-06 & 3.89E-06 \nl
\enddata
\tablenotetext{a}{Only mass fractions of stable or long-lived nuclides in access of $X\approx5\times10^{-7}$ are listed here. For $^{26}$Al the listed values refer to the ground state. The labels ``xNe/C", ``C/Ne" and ``H" refer to explosive Ne/C burning, convective shell C/Ne burning, and convective core H burning, respectively.}
\tablenotetext{b}{Initial mass fractions at beginning of post-processing calculations (from Limongi \& Chieffi 2006).}
\tablenotetext{c}{Final mass fractions at the end of post-processing calculations, obtained from the ``standard" calculations (i.e., without any reaction rate variations).}
\tablenotetext{d}{Solar system mass fractions, for comparison (from Lodders 2003).}
\end{deluxetable}
Our strategy regarding which and how many reaction rates to vary was as follows. We started by considering the net abundance flows (i.e, the difference of total abundance flows between a given forward and corresponding reverse reaction), integrated over the entire duration of a ``standard" post-processing calculation. These flows will be displayed graphically below for each of the three burning regimes. All rates of reactions with net flows within 3 orders-of-magnitude of the maximum flow were then selected for the variation procedure. Obviously, the forward and corresponding reverse reaction rate need to be multiplied by the same variation factor. We added to this list all reactions that either destroy or produce $^{26}$Al, $^{27}$Al and $^{25}$Mg, if these were not taken into account already. Furthermore, a number of selected other reactions, for example, $^{12}$C($^{12}$C,n)$^{23}$Mg and $^{24}$Mg(p,$\gamma$)$^{25}$Al, were added to the list. We find it unlikely that any other reaction not identified by the above procedure has a major impact on $^{26}$Al nucleosynthesis in massive stars.

A number of important issues need to be considered in detail when performing any reaction rate sensitivity study using post-processing calculations. First, it is assumed that the nuclear reaction rates to be varied do not impact the nuclear energy generation. If a given reaction rate variation changes both the energy generation and the final $^{26}$Al abundance, then this result has no obvious meaning. Clearly, in such cases the rate should be varied using the full, self-consistent stellar model. For example, changing the rate of the $^{20}$Ne($\gamma$,$\alpha$)$^{16}$O reaction, the process that initiates Ne burning, influences the $^{26}$Al abundance, although the effect is small, as will be seen below. Throughout this work, we carefully checked that a rate variation did not impact at the same time the energy generation. Second, it is very important to verify that a given temperature-density-time evolution is followed {\it precisely} in a post-processing calculation. The time step is numerically adjusted to track the abundance evolutions above some limiting abundance value. For example, spurious abundance variations may occur if a time step misses the peak temperature even by a few percent. Therefore, we carefully checked that the temperature-density evolution is followed closely (within a fraction of a percent). 

Finally, we need to address the issue of stellar versus laboratory $\beta$-decay rates. The stellar plasma affects $\beta$-decays in a number of ways. First, at high temperatures thermally excited states in the $\beta$-decaying nuclide may undergo transitions to levels in the daughter nuclide. Second, at high (electron) densities the decay constants for electron (or positron) capture will increase. Both effects generally cause a change in the total $\beta$-decay rate. Many previous stellar model studies  employed the stellar $\beta$-decay constants calculated by Fuller, Fowler \& Newman (1982) for the mass range of A=21-60, or the more modern results of Oda et al. (1994), that are based on shell model calculations, for the mass range of A=17-39. For the purposes of the present work, the highest temperature and density values are encountered in explosive Ne/C burning, with peak values of $T=2.3$ GK and $\rho=3.2\times10^5$ g/cm$^{3}$ (see below). By inspecting the tables of Oda et al. (1994) in the region A$\leq$30, we find that for the temperatures and densities of interest here the stellar and laboratory $\beta$-decay rates are very similar in magnitude. The only exceptions are the $\beta$-decays of the long-lived species $^{22}$Na and $^{24}$Na. However, their destruction via the processes ($\gamma$,n), ($\gamma$,p) or (p,$\gamma$) is much faster compared to the $\beta$-decays and, therefore, their stellar $\beta$-decay rate is unimportant at high values of $T$ and $\rho$. In conclusion, it is sufficient to adopt laboratory $\beta$-decay rates throughout this work, except for the $\beta$-decay of (thermalized) $^{26}$Al$^t$. For the calculation of this decay we only take the ground and isomeric state into account. The decay constants are listed in Appendix \ref{appdec} and agree with the more comprehensive results of Oda et al. (1994) for temperatures and densities below 5 GK and 10$^6$ g/cm$^3$, respectively.

\subsection{Explosive Ne/C burning}
\subsubsection{Standard calculation}
\begin{figure}
\includegraphics[scale=.45]{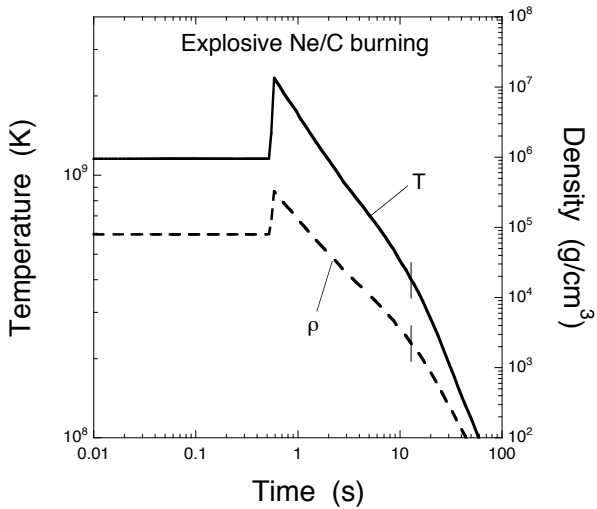}
\caption{Temperature-density-time evolution for explosive Ne/C burning. The profile was obtained from a hydrodynamic model of a 20$M_\odot$ star (Limongi \& Chieffi 2006). In the present post-processing calculation, the evolution was only followed until $t=12.8$ s (vertical lines), since for later times $T$ and $\rho$ decline to values where no additional $^{26}$Al synthesis is occurring. The peak temperature and density, near $t=0.6$ s, amount to $T=2.3$ GK and $\rho=3.2\times10^5$ g/cm$^{3}$, respectively (see text). \label{figxNetrho}}
\end{figure}

At the beginning of the burning, the most abundant nuclides are (in order) $^{16}$O, $^{20}$Ne, $^{24}$Mg, $^{28}$Si and $^{12}$C (Tab. \ref{tababun}). The temperature-density-time profile for simulating explosive Ne/C burning is shown in Fig. \ref{figxNetrho}. It has been extracted from a stellar model calculation of a 20$M_\odot$ star ($\S$ \ref{starmodel}). Specifically, we select a mass coordinate of 2.04$M_\odot$, corresponding to the zone where the maximum abundance of $^{26}$Al is produced during the explosion. Temperature and density peak at $T=2.3$ GK and $\rho=3.2\times10^5$ g/cm$^3$, respectively. The evolution is followed in a post-processing simulation over a total time of $t=12.8$ s. At this point the temperature has declined to $T=0.4$ GK and no additional $^{26}$Al synthesis is occurring. We assume at this stage thermal equilibrium for $^{26}$Al, i.e., the network contains only a single species, $^{26}$Al$^t$.

\begin{figure*}
\includegraphics[scale=1.0]{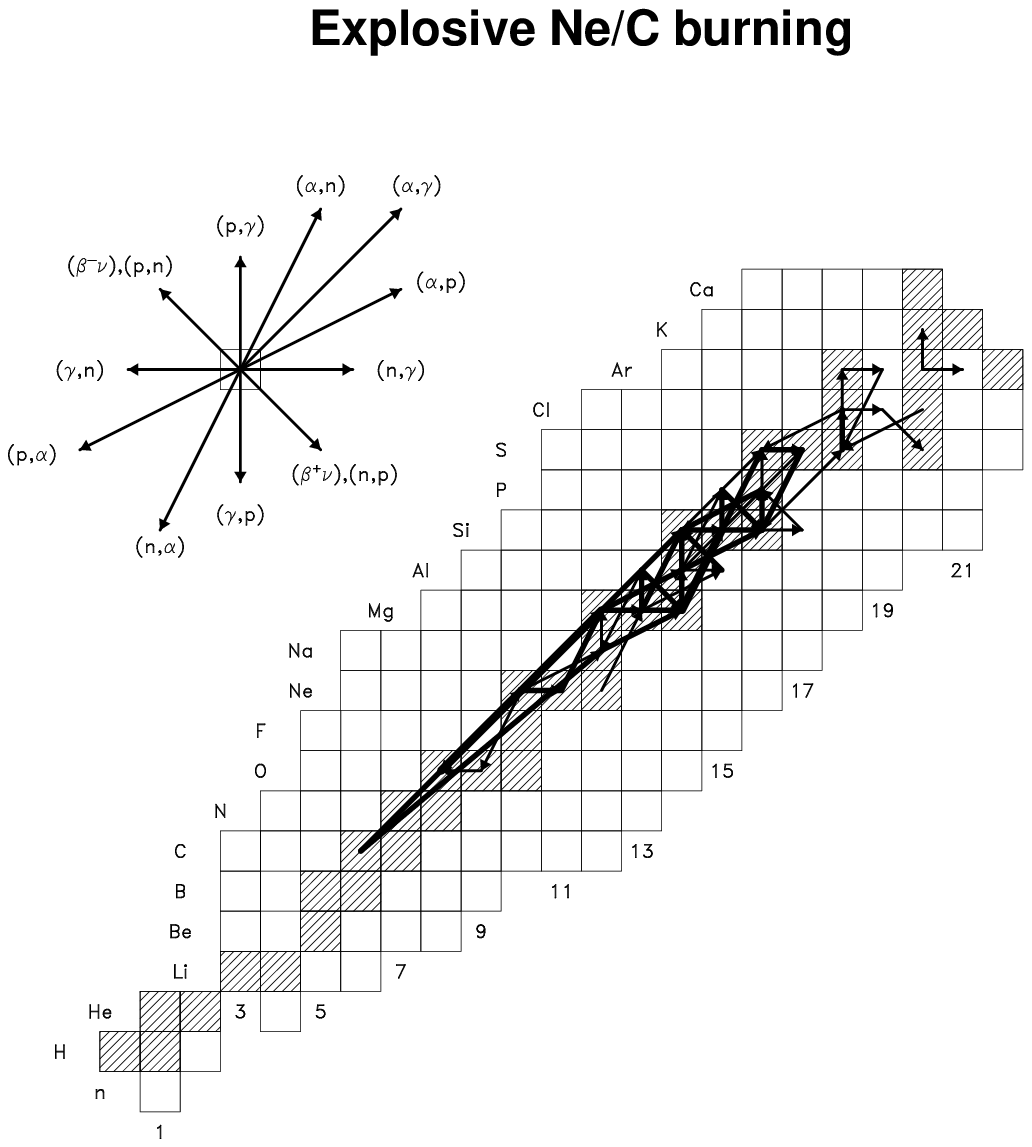}
\caption{Net abundance flows, obtained for a post-processing network calculation of explosive Ne/C burning, integrated over a total running time of $t=12.8$ s. The $T$-$\rho$ profile for this simulation is shown in Fig. \ref{figxNetrho}. The network consists of all nuclides shown as squares. The strongest net abundance flows, i.e., those within one, two, and three orders of magnitude of the maximum flow, are displayed by the thickest arrows, arrows of intermediate thickness, and the thinnest arrows, respectively. Thermal equilibrium for $^{26}$Al has been assumed (i.e., the network contains only a single species, $^{26}$Al$^t$).\label{figxNeflow}}
\end{figure*}

\begin{figure}
\includegraphics[scale=.49]{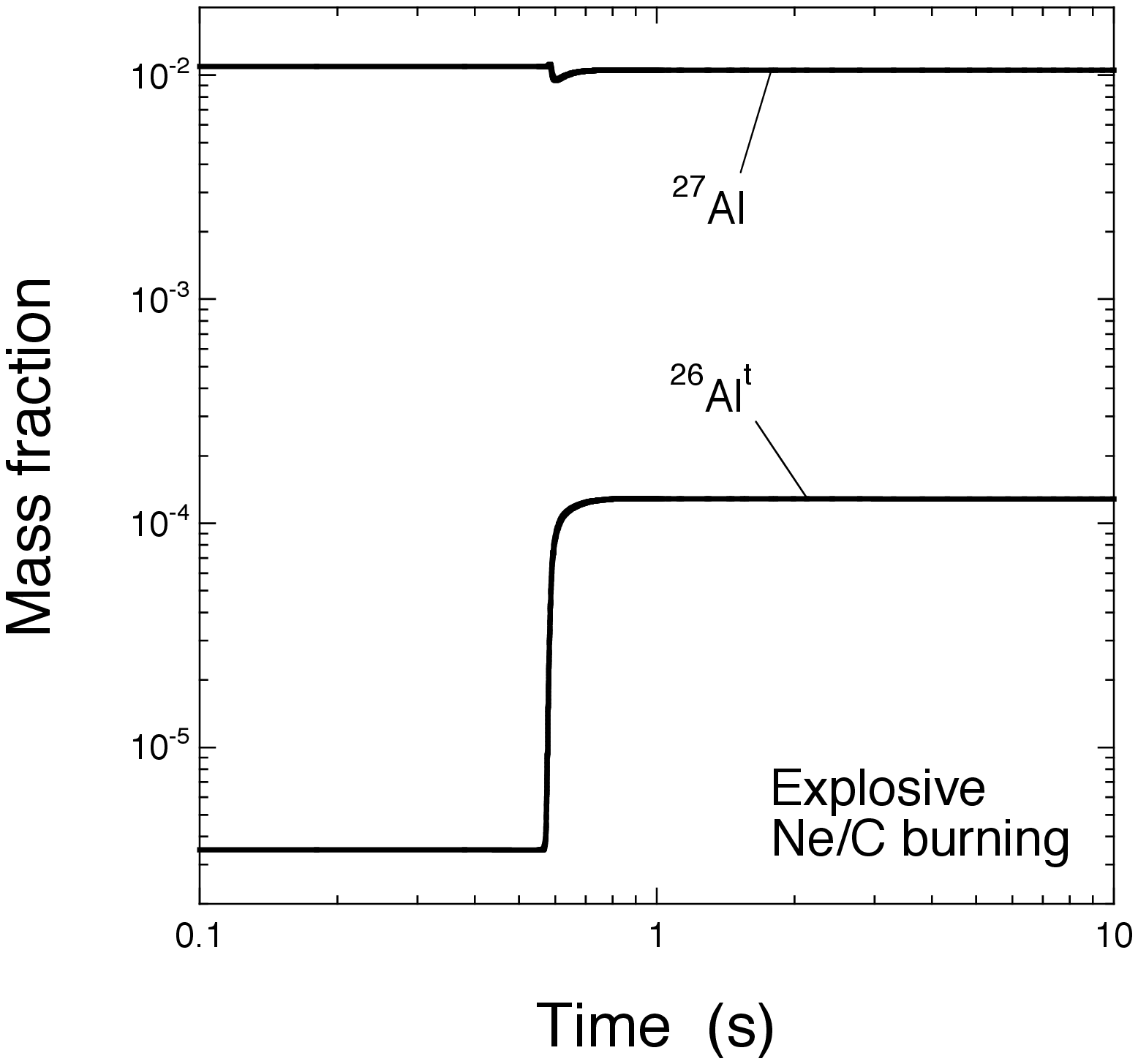}
\caption{Abundance evolution (by mass) of $^{26}$Al$^t$ and $^{27}$Al during explosive Ne/C burning. The $T$-$\rho$ profile for this post-processing network simulation is shown in Fig. \ref{figxNetrho}. The calculation assumes a single species of (thermalized) $^{26}$Al.
\label{figxNealal}}
\end{figure}

The net abundance flows, integrated over a total running time of $t=12.8$ s, for the ``standard" calculation are displayed in Fig. \ref{figxNeflow}. They provide a first impression regarding the nucleosynthesis and indicate the degree of ``nuclear activity". The network consists of all nuclides shown as squares. The strongest net abundance flows, i.e., those within one, two, and three orders of magnitude of the maximum flow, are displayed by the thickest arrows, arrows of intermediate thickness, and the thinnest arrows, respectively. The strongest net flows belong to the reactions $^{20}$Ne($\gamma$,$\alpha$)$^{16}$O and $^{20}$Ne($\alpha$,$\gamma$)$^{24}$Mg that drive explosive Ne burning. The released $\alpha$-particles induce a network of secondary reactions, giving rise to a small, but significant, abundance of light particles. During the explosive burning the mass fractions of protons, $\alpha$-particles and neutrons reach maximum values of 1.0$\times$10$^{-8}$, 1.1$\times$10$^{-5}$ and 3.8$\times$10$^{-11}$, respectively. The main direct process of $^{26}$Al$^t$ synthesis, in terms of the net abundance flow, is $^{25}$Mg(p,$\gamma$)$^{26}$Al$^t$, with $^{25}$Mg produced by the $^{24}$Mg(n,$\gamma$)$^{25}$Mg reaction. The main neutron sources are $^{25}$Mg($\alpha$,n)$^{28}$Si and $^{26}$Mg($\alpha$,n)$^{29}$Si. On the other hand, $^{26}$Al$^t$ is predominantly destroyed via $^{26}$Al$^t$(n,p)$^{26}$Mg and, to a lesser degree, $^{26}$Al$^t$(n,$\alpha$)$^{23}$Na. In particular, the $^{26}$Al$^t$(p,$\gamma$)$^{27}$Si reaction is entirely negligible under explosive burning conditions. These general features have already been discussed by Limongi \& Chieffi (2006). The abundance evolutions of the species $^{26}$Al$^t$ and $^{27}$Al are shown in Fig. \ref{figxNealal}. While the abundance of the latter nuclide is approximately constant throughout the calculation, the abundance of the former species increases by more than an order of magnitude during the explosion. The abundance ratio, $X(^{26}Al^t)/X(^{27}Al)$, increases from an initial value of $3.2\times10^{-4}$ to a final value of $1.2\times10^{-2}$ (see Tab. \ref{tababun}).

\subsubsection{Reaction rate variations\label{xNevary}}
Subsequently, the rates of 70 pairs of forward and reverse reactions were varied. Those reactions whose rate changes have the strongest effect on the final $^{26}$Al yield (i.e., at time $t=12.8$ s) are listed in Tab. \ref{tabxNe}. All other rate changes, as well as those labeled by ``..." in the table, produced $^{26}$Al$^t$ abundance changes of less than 20\%. The reactions are listed in approximate order of importance, as measured by their impact on the final $^{26}$Al abundance. The last two columns display the source of the rate and the reported rate uncertainty at a temperature near the peak of the explosion (i.e., where most of the nucleosynthesis is occurring). Disregarding for a moment the actual rate uncertainties, the six reactions with the strongest impact on $^{26}$Al$^t$ nucleosynthesis are: $^{26}$Al$^t$(n,p)$^{26}$Mg, $^{25}$Mg(p,$\gamma$)$^{26}$Al$^t$, $^{25}$Mg($\alpha$,n)$^{28}$Si, $^{24}$Mg(n,$\gamma$)$^{25}$Mg,  $^{20}$Ne($\alpha$,$\gamma$)$^{24}$Mg, and $^{30}$Si(p,$\gamma$)$^{31}$P. The first and second reaction destroys and produces $^{26}$Al$^t$, respectively, while the third and fourth reaction destroys and produces, respectively, the $^{25}$Mg seed. The fifth reaction produces $^{24}$Mg, from which $^{25}$Mg is synthesized via neutron capture. The sensitivity of the final $^{26}$Al$^t$ abundance to any of these rate changes is thus not surprising.

\begin{deluxetable}{llllllc}
\tablewidth{0pt}
\tablecaption{Factor changes of final $^{26}$Al$^t$ abundance resulting from reaction rate variations for explosive Ne/C burning\tablenotemark{a}, assuming thermal equilibrium for $^{26}$Al\label{tabxNe}}
\tablehead{
\colhead{Reaction\tablenotemark{b}} &  \multicolumn{4}{c}{Rate multiplied by}  \nl 
\cline{2-5}
&  \colhead{10} & \colhead{2} & \colhead{0.5} & \colhead{0.1}   & \colhead{Source\tablenotemark{c}}    &  \colhead{Uncertainty\tablenotemark{d}}   
}
\startdata
$^{25}$Mg($\alpha$,n)$^{28}$Si		      &  0.10  &  0.49       &  1.8         &  4.0         &  nacr   &  18\% \\
$^{24}$Mg(n,$\gamma$)$^{25}$Mg		      &  5.2    &  1.6         &  0.61       &  0.24       &  ka02  &   \\
$^{26}$Al$^t$(n,p)$^{26}$Mg			      &  0.14  &  0.58       &  1.6         &  3.2         &  present   &   \\
$^{25}$Mg(p,$\gamma$)$^{26}$Al$^t$          &  1.7    &  1.4          &  0.58      &  0.14       &  il10  &  4\% \\
$^{30}$Si(p,$\gamma$)$^{31}$P		      &  0.51  &  0.77       &  1.3         &  2.0         &  il10  &  14\% \\
$^{20}$Ne($\alpha$,$\gamma$)$^{24}$Mg   &  1.8    &  1.4         &  0.64       &  0.28       &  il10  &  11\% \\
$^{27}$Al($\alpha$,p)$^{30}$Si	              &  1.5    &  \nodata  &  \nodata  &  0.72       &  rath  &   \\
$^{29}$Si($\alpha$,n)$^{32}$S		              &  0.65  &  \nodata  &  \nodata  &  1.3         &  rath  &   \\
$^{24}$Mg($\alpha$,$\gamma$)$^{28}$Si    &  0.62  &  \nodata  &  \nodata  &  \nodata  &  il10   &  6\% \\
$^{24}$Mg($\alpha$,p)$^{27}$Al                   & \nodata&  \nodata  &  \nodata  &  0.65      &  il10   &  6\% \\
$^{27}$Al(p,$\gamma$)$^{28}$Si                  &  0.60  &  \nodata  &  \nodata  &  \nodata  &  il10  &  3\% \\
$^{25}$Mg($\alpha$,p)$^{28}$Al		      &  0.59  &  \nodata  &  \nodata  &  \nodata  &  rath  &   \\
$^{26}$Al$^t$(n,$\alpha$)$^{23}$Na	      &  0.55  &  \nodata  &  \nodata  &  \nodata  &  present  &   \\
$^{25}$Mg(n,$\gamma$)$^{26}$Mg              &  0.75  &  \nodata  &  \nodata  &  \nodata  &  ka02  &   \\
$^{28}$Si(n,$\gamma$)$^{29}$Si                  &  1.4    &  \nodata  &  \nodata  &  \nodata  &  ka02  &   \\
$^{29}$Si(p,$\gamma$)$^{30}$P                   &  1.4    &  \nodata  &  \nodata  &  \nodata  &  il10  &  7\% \\
$^{32}$S(n,$\gamma$)$^{33}$S                    &  1.2    &  \nodata  &   \nodata  &  \nodata  &  ka02  &   \\
$^{26}$Al$^t$($\alpha$,p)$^{29}$Si               &  0.72  &  \nodata  &  \nodata  &  \nodata  &  rath  &   \\
$^{26}$Mg(p,$\gamma$)$^{27}$Al                &  1.2    &  \nodata  &  \nodata  &  \nodata  &  il10  &  4\% \\
\enddata
\tablenotetext{a}{The temperature-density-time profile is extracted from a hydrodynamic model of a 20$M_\odot$ star of initial solar metallicity, see Limongi \& Chieffi (2006).}
\tablenotetext{b}{In total, the rates of 70 different reactions were varied. Listed are only those reactions whose rate changes have the strongest effect on the $^{26}$Al$^t$ yield. All other rate changes, as well as those labeled by ``...", produced abundance changes of less than 20\%. The reactions are listed in approximate order of importance. Thermal equilibrium for $^{26}$Al has been assumed, i.e., the network contains only a single species, $^{26}$Al$^t$.}
\tablenotetext{c}{Reaction rate references: (nacr) Angulo et al. 1999 (NACRE); (ka02) Dillmann et al. (2006) (KADoNiS v0.2); (rath) Rauscher \& Thielemann (2000); (il10) Iliadis et al. (2010); (present) hybrid rates, see Appendix \ref{al26na} and \ref{al26np}.}
\tablenotetext{d}{Reaction rate uncertainty near a temperature of 2.3 GK, at the peak of the explosion; no entry implies that the rate uncertainty is difficult to quantify (see text).}
\end{deluxetable}

The manner by which the sixth reaction impacts the synthesis of $^{26}$Al$^t$ is interesting. In fact, the sequence $^{30}$Si(p,$\gamma$)$^{31}$P(p,$\alpha$)$^{28}$Si is the main consumer of free protons (together with the proton captures on $^{26}$Mg and $^{27}$Al). When the rate of the  $^{30}$Si(p,$\gamma$)$^{31}$P reaction is reduced by an order of magnitude, the number of available protons increases near the peak of the explosion and, consequently, more $^{25}$Mg nuclei are converted to $^{26}$Al$^t$. There are 13 more reactions listed in Tab. \ref{tabxNe} and their mechanisms by which they impact the final $^{26}$Al$^t$ abundance can be easily deduced from arguments similar to those given above. The only reaction that we found to influence somewhat the $^{26}$Al$^t$ yield but is not listed in Tab. \ref{tabxNe} is $^{20}$Ne($\gamma$,$\alpha$)$^{16}$O. Varying this rate by a factor of 10 changes the $^{26}$Al$^t$ abundance by a factor of $\approx$2. However, the estimated rate uncertainty of this reaction amounts to only 13\%
(Iliadis et al. 2010) and thus the actual effect is relatively small.

\subsubsection{Reaction rate uncertainties\label{uncert}}
Before proceeding, notice the sources of our reaction rates, listed in column 6 of Tab. \ref{tabxNe}. Of the 19 reactions listed, the rates of: (i) 8 reactions are available from the Monte Carlo procedure (Iliadis et al. 2010; $\S$ \ref{physlib}); (ii) 6 reactions are adopted from the statistical model (Rauscher \& Thielemann 2000); (iii) 4 reactions are obtained from KADoNiS v0.2 (Dillmann et al. 2006); and (iv) only one is adopted from NACRE (Angulo et al. 1999). Note that none of these rates rely anymore on outdated information from Caughlan \& Fowler (1988).

We now turn to a discussion of reaction rate uncertainties. These are listed for a temperature of $T=2.5$ GK, near the peak of the explosion, in the last column of Tab. \ref{tabxNe}, when reported in the original source. Rate uncertainty estimates are of obvious importance. Suppose a rate variation of a particular reaction by a factor of 10 changes the $^{26}$Al$^t$ abundance by the same factor. Then one may conclude that this particular reaction rate should be known with rather small uncertainty. On the other hand, if a particular rate variation barely affects the abundance of $^{26}$Al$^t$, one may tolerate a much larger uncertainty. In reality, however, the issue is much more complicated and one is usually confronted with the following questions when considering rate uncertainties reported in the literature. What is the (statistical) meaning of a reported rate uncertainty? Is a presumed experimental rate at a given temperature directly based on data, or is it based on a normalization of (theoretical) Hauser-Feshbach rates? Even if a rate is directly based on data, how large is the stellar enhancement factor that must usually be obtained from Hauser-Feshbach models? What may one estimate for a rate uncertainty if no values are reported in the literature? And, even if a given rate is directly based on data and if the stellar enhancement factor is negligible at a given temperature, are there possible systematic errors that were not taken into account in the reported rate uncertainty? All of these issues play an important role and thus reported rate uncertainties are frequently difficult to assess. Below we will give a few examples to emphasize these points. 

The first reaction listed in Tab. \ref{tabxNe}, $^{25}$Mg($\alpha$,n)$^{28}$Si, strongly affects the final $^{26}$Al$^t$ abundance. Varying the rate by a factor of 10 (2) changes the $^{26}$Al$^t$ yield by a factor of 0.1 (0.5). The rate is adopted from the NACRE compilation (Angulo et al. 1999), and its reported uncertainty of $\approx$18\% near $T=2.5$ GK may on first sight indicate a rather reliable rate. However, not enough information is provided in Angulo et al. (1999) to understand how exactly this value of uncertainty has been obtained. Also, beyond $T=2$ GK, i.e., the highest temperature for which the rate is directly based on data, the rate was extrapolated with the aid of (theoretical) Hauser-Feshbach model results. Furthermore, even at the lower temperatures, the rate seems to be based on data from an unpublished thesis. Considering these arguments together with the importance of the $^{25}$Mg($\alpha$,n)$^{28}$Si reaction, there is no doubt in our minds that this particular reaction should be a target of future experimental work. Consider, on the other hand the fourth reaction listed in Tab. \ref{tabxNe}, $^{25}$Mg(p,$\gamma$)$^{26}$Al$^t$. Varying the rate by a factor of 0.1 (0.5) changes the $^{26}$Al$^t$ yield by a factor of 0.14 (0.58). A rate uncertainty of only 4\% near $T=2.5$ GK has been reported by Iliadis et al. (2010). This value has been obtained from a Monte Carlo procedure, implying a statistically meaningful probability coverage (68\%). The rate near the peak of the explosion is directly based on data, i.e., no extrapolation using theoretical reaction models is needed. Furthermore, the experimental rate is normalized to a well-known standard resonance strength (for details, see Iliadis 2007). In conclusion, at present there is less compelling reason for remeasuring this reaction at higher energies compared to the previous case. We emphasize again that each reaction must be treated as a special case and that a reported rate uncertainty needs to be considered carefully. For readers interested in the present status of specific reactions, we provide brief discussions in Appendix \ref{specrates}.  

The set of rates shown in Tab. \ref{tabxNe} that are based on Hauser-Feshbach theory (labelled by ``rath") represent a special case. It has been claimed by Rauscher \& Thielemann (2000) that ``...the accuracy of the rates is estimated to be within a factor of 1.5-2...". Obviously, if too few resonances contribute to the rate at a given temperature, the statistical model will provide a poor description. For this reason, Rauscher \& Thielemann (2000) provide a minimum temperature estimate below which the Hauser-Feshbach rates become inaccurate. This minimum temperature value is calculated from a parameterization of nuclear level densities, assuming that at least 10 levels (Rauscher, Thielemann \& Kratz 1997) are located in the astrophysically important energy window (e.g., the Gamow peak for charged-particle reactions). Note that for all of the reactions labeled ``rath" in Tab. \ref{tabxNe} the peak temperature of the explosion ($T=2.3$ GK) far exceeds the minimum temperature required for the applicability of the Hauser-Feshbach model according to Rauscher \& Thielemann (2000). Unfortunately, the above claims are not supported when comparing the Hauser-Feshbach rates with results that are directly based on experiment. The issue was discussed in Iliadis et al. (2001), who found that for several reactions involving A=20-40 mass targets ``... the deviation between theoretical and experimental rates far exceeds the usually quoted factor of 2 reliability of statistical model results ...". Clearly, more work is required to resolve this controversy. At this point it may be argued that all of the reactions labeled by ``rath" in Tab. \ref{tabxNe} should be targets for future experimental work, including the important destruction reactions $^{26}$Al$^t$(n,p)$^{26}$Mg and $^{26}$Al$^t$(n,$\alpha$)$^{23}$Na (labeled ``present"; see Appendix \ref{al26np} and \ref{al26na}).

Neutron capture rates represent another special case. We adopted for these the results presented in the KADoNiS v0.2 evaluation (Dillmann et al. 2006; these rates are labelled by ``ka02" in Tab. \ref{tabxNe}). The most important neutron capture reaction for the purposes of the present work is $^{24}$Mg(n,$\gamma$)$^{25}$Mg, as is apparent from the table. In order to obtain a better sense for the uncertainties, we will briefly discuss how the KADoNiS evaluated rates have been obtained and what information is actually incorporated in reaction rate libraries. The arguments below apply equally to the other reactions listed in Tab. \ref{tabxNe}, i.e., $^{25}$Mg(n,$\gamma$)$^{26}$Mg, $^{28}$Si(n,$\gamma$)$^{29}$Si and $^{32}$S(n,$\gamma$)$^{33}$S. The KADoNiS evaluation tabulates recommended rates for the range of $kT=5-100$ keV (corresponding to $T=0.06-1.2$ GK). For the neutron captures on $^{24,25}$Mg, $^{28}$Si and $^{32}$S the rates are obtained from experimental data on resonance properties (with some theoretical corrections for direct neutron capture contributions, if applicable) over the entire tabulated temperature range. According to the KADoNiS evaluation, the ``relative uncertainties [of the rates] are similar to those quoted for the 30 keV data" (12\% for neutron capture on $^{24}$Mg). The tabulated rates include the stellar enhancement factor ($\S$ \ref{physlib}), although these are predicted to be close to unity for the (n,$\gamma$) reactions mentioned above. Note that for explosive Ne/C burning the rates are needed at temperatures ($T\approx2.3$ GK) that have not been covered by experiments. Thus it is not obvious how to extrapolate the rates from lower temperatures, where they are based on experimental data, to much higher temperatures. Furthermore, it must be pointed out that the experimental KADoNiS rates are not directly used in reaction rate libraries (including ours). What is usually incorporated for neutron captures are Hauser-Feshbach rates, which are normalized to the experimental rates at a single temperature ($kT=30$ keV or $T=0.35$ GK). Since the level density for targets in the mass $A\leq40$ range at $kT=30$ keV may be too small for the application of statistical models, an additional systematic uncertainty is introduced when extrapolating such normalized rates to higher temperatures. For example, in the case of $^{24}$Mg(n,$\gamma$)$^{25}$Mg the (experimental) KADoNiS rate at the upper temperature cutoff ($T=1.2$ GK) deviates from the {\it normalized} Hauser-Feshbach rate already by $\approx$40\%. Considering the above arguments, we estimate a rate uncertainty of a factor of $2-3$ for the neutron captures on $^{25,25}$Mg, $^{28}$Si and $^{32}$S near the peak of explosive Ne/C burning. Recall from Tab. \ref{tabxNe} that varying the $^{24}$Mg(n,$\gamma$)$^{25}$Mg rate by a factor of 2 increases the $^{26}$Al$^t$ yield by a factor of 1.6. Clearly, a more reliable experimental rate for $^{24}$Mg(n,$\gamma$)$^{25}$Mg at higher temperatures is urgently needed.

\subsubsection{Thermal equilibration\label{secthermxNe}}
We will now consider the issue of thermal equilibration. Recall that we assumed so far a single species of $^{26}$Al, implying thermal equilibrium ($^{26}$Al$^t$). We will now relax this assumption and follow the equilibration numerically in the network calculation. To this end, we introduce five different species of $^{26}$Al, as explained in $\S$ \ref{thermequi}. The required $\gamma$- and $\beta$-decay transitions between and from these levels are discussed in detail in Appendix \ref{appdec}. The price we pay is that additional reaction rates, involving $^{26}$Al$^g$ and $^{26}$Al$^m$ separately, have to be incorporated into the network (see Appendix \ref{appreac}). As will be seen, some of these rates are highly uncertain.

\begin{figure}
\includegraphics[scale=.55]{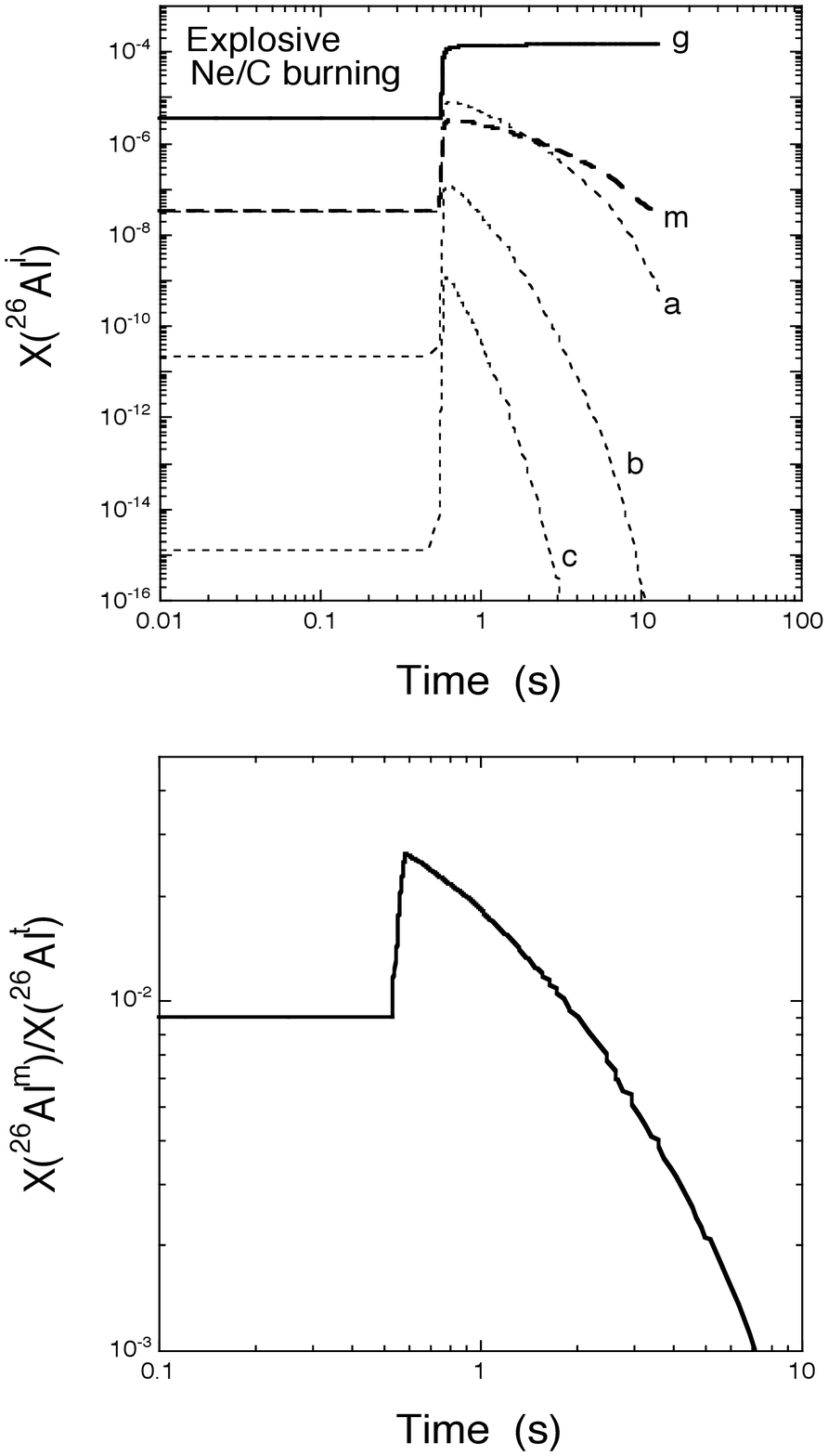}
\caption{Abundance evolution (by mass) of $^{26}$Al during explosive Ne/C burning. The $T$-$\rho$ profile for this post-processing network simulation is shown in Fig. \ref{figxNetrho}. The calculation assumes five species of $^{26}$Al: ground state ($g$), isomeric state ($m$), and three excited levels ($a$, $b$, $c$); see Fig. \ref{figlevel}. The communication of the different $^{26}$Al levels via $\gamma$-ray transitions is explicitly taken into account. (Top) Abundance evolution of different $^{26}$Al species; (Bottom) Numerically simulated fraction of total $^{26}$Al abundance that resides in the isomeric state. The curve is indistinguishable from the one calculated assuming a Boltzmann distribution (i.e., thermal equilibrium).
\label{figxNe5s}}
\end{figure}

As a first step, we performed a standard post-processing network calculation (with recommended rates) using the same temperature-density-time evolution as before (Fig. \ref{figxNetrho}). The final $^{26}$Al$^{g}$ abundance, at $t=12.8$ s, is found to be identical to our earlier result obtained assuming thermal equilibrium (Tab. \ref{tababun}). Thus the latter assumption seems to be justified. An impression can be gained from Fig. \ref{figxNe5s}, showing the abundance evolution of $^{26}$Al levels. The top part displays the abundances for individual $^{26}$Al species and it is apparent that at any given time the $^{26}$Al$^g$ abundance dominates over those of the other species. The bottom part displays the fraction of the total $^{26}$Al abundance that resides in the isomeric state. This curve is directly obtained from the network calculation, but is indistinguishable from the one calculated assuming a Boltzmann distribution (i.e., thermal equilibrium).

Next, the rates of 20 different reactions and transitions, together with their inverse rates, were varied individually by factors of 100, 10, 2, 0.5, 0.1 and 0.01. This list contained all nuclear reactions that produced and destroyed $^{26}$Al$^g$ or $^{26}$Al$^m$. It also included those $\beta$- and $\gamma$-ray decay rates of $^{26}$Al$^x$ levels that were estimated using the shell model, as described in $\S$ \ref{thermequi} and shown in Fig. \ref{figlevel}. Experimentally obtained $\beta$- and $\gamma$-ray decay rates have not been varied since their uncertainties are very small. The final $^{26}$Al$^g$ abundance, after each rate variation, was then compared to the standard calculation. The results are listed in Tab. \ref{tabxNe5s}. It can be seen that of the 20 interactions only eight, all of them reactions, influence the final $^{26}$Al$^g$ yield. In other words, even a variation by a factor of 100 of the shell-model based $\beta$- or $\gamma$-ray decay rates seems to have no effect on the $^{26}$Al$^g$ abundance. The eight reactions displayed in Tab. \ref{tabxNe5s} are listed in approximate order of importance. It must be noted that the impact of these rates, with one exception, on the final abundance of $^{26}$Al$^g$ is moderate. For example, consider the second and third reaction, $^{25}$Mg(p,$\gamma$)$^{26}$Al$^g$ and $^{25}$Mg(p,$\gamma$)$^{26}$Al$^m$. Even a rather small rate variation (by a factor of 2) influences the final $^{26}$Al yield (by $\approx$30\%). However, these rates are based on experimental information (Iliadis et al. 2010) and their Monte Carlo uncertainties are predicted to amount to only 4-5\% near the peak of the explosion. 

\begin{deluxetable}{llllllllc}
\tablewidth{0pt}
\tablecaption{Factor changes of final $^{26}$Al$^g$ abundance resulting from reaction rate variations for explosive Ne/C burning\tablenotemark{a}, assuming five species of $^{26}$Al\label{tabxNe5s}}
\tablehead{
\colhead{Reaction\tablenotemark{b}} &  \multicolumn{6}{c}{Rate multiplied by}  \nl 
\cline{2-7}
&  \colhead{100} & \colhead{10} & \colhead{2} & \colhead{0.5} & \colhead{0.1} & \colhead{0.01}  & \colhead{Source\tablenotemark{c}}    &  \colhead{Uncertainty\tablenotemark{d}}   
}
\startdata
$^{26}$Al$^g$(n,p)$^{26}$Mg                    &  0.017  & 0.14       &  0.57         & 1.6         & 2.9          & 3.8          &  present   &    \\
$^{25}$Mg(p,$\gamma$)$^{26}$Al$^g$     &  1.7      & 1.6         &  1.3           & 0.71       & 0.45        & 0.38         &  il10    &   4\% \\
$^{25}$Mg(p,$\gamma$)$^{26}$Al$^m$    &  1.6      & 1.6         &  \nodata    & \nodata  & 0.79        & 0.79        &  il10      &   5\% \\
$^{26}$Al$^g$($\alpha$,p)$^{29}$Si          &  0.21     & 0.71      &  \nodata    & \nodata  &  \nodata  &  \nodata   &  rath   &    \\
$^{26}$Al$^g$(n,$\alpha$)$^{23}$Na         &  0.21     & 0.54      &  \nodata    & \nodata  &  \nodata  &  \nodata   &  present   &   \\
$^{26}$Al$^m$(n,p)$^{26}$Mg                   &  0.36     & \nodata  &  \nodata    & \nodata  &  \nodata  &  \nodata   &  present   &    \\
$^{26}$Al$^g$(p,$\gamma$)$^{27}$Si       &  0.52     & \nodata  &  \nodata    & \nodata  &  \nodata  &  \nodata   &  il10      &  7\% \\
$^{26}$Al$^m$(n,$\alpha$)$^{23}$Na         &  0.79    & \nodata  &  \nodata    & \nodata  &  \nodata  &  \nodata   &  present   &    \\
\enddata
\tablenotetext{a}{The temperature-density-time profile is extracted from a hydrodynamic model of a 20$M_\odot$ star of initial solar metallicity, see Limongi \& Chieffi (2006).}
\tablenotetext{b}{In total, the rates of 20 different reactions producing or destroying $^{26}$Al$^g$ and $^{26}$Al$^m$ were varied. Listed are only those reactions whose rate changes have the strongest effect on the $^{26}$Al$^g$ yield. All other rate changes, as well as those labeled by ``...", produced abundance changes of less than 20\%. The reactions are listed in approximate order of importance. No thermal equilibrium for $^{26}$Al has been explicitly assumed, i.e., the network contains five different species ($^{26}$Al$^g$, $^{26}$Al$^m$, $^{26}$Al$^a$, $^{26}$Al$^b$, $^{26}$Al$^c$) and takes the interactions between them into account.}
\tablenotetext{c}{Reaction rate references: (il10) Iliadis et al. 2010; (rath) Rauscher \& Thielemann  2000; (present) hybrid rate, see Appendix \ref{al26na} and \ref{al26np}. In the latter three cases, we assumed that the rate involving $^{26}$Al$^g$ or $^{26}$Al$^m$ is the same as the rate for $^{26}$Al$^t$ (see comments in Appendix \ref{appreac}).}
\tablenotetext{d}{Reaction rate uncertainty near a temperature of 2.3 GK, at the peak of the explosion; no entry implies that the rate uncertainty is difficult to quantify.}
\end{deluxetable}

The one exception is the $^{26}$Al$^g$(n,p)$^{26}$Mg reaction rate. Increasing this rate by a factor of 100 changes the $^{26}$Al$^g$ abundance by a factor of 0.017. We adopted for this reaction the same rates as for $^{26}$Al$^t$(n,p)$^{26}$Mg (see Appendix \ref{appreac}). Our numerical results indicate that  even a factor of 100 variation in the $^{26}$Al$^g$(n,p)$^{26}$Mg rate does not change the thermal equilibrium abundance ratio of $^{26}$Al$^m$ and $^{26}$Al$^g$ (Fig. \ref{figxNe5s}). In fact, comparison of the first entry of Tab. \ref{tabxNe5s} with the third entry of Tab. \ref{tabxNe} immediately reveals that the $^{26}$Al(n,p)$^{26}$Mg reaction impacts the final $^{26}$Al abundance by the same factor changes, no matter if a single (thermalized) or five species of $^{26}$Al are used in the simulation. In conclusion, $^{26}$Al is in thermal equilibrium\footnote{The reader may suspect circular reasoning in our arguments, in the sense that we used the same rates for the ground and isomeric states as for the thermalized $^{26}$Al target, and then conclude that $^{26}$Al is in thermal equilibrium. However, this is not the case since our assumption for the nominal rates serves as a starting point only and we fully explore individual rate changes by factors up to 100. For more information, see App.~\ref{appreac}.} during explosive Ne/C burning and, consequently, there is no need to introduce the extra complication of five $^{26}$Al levels, and their mutual interactions, into the reaction network.

\subsection{Convective shell C/Ne burning}
\subsubsection{Standard calculation}
\begin{figure}
\includegraphics[scale=.45]{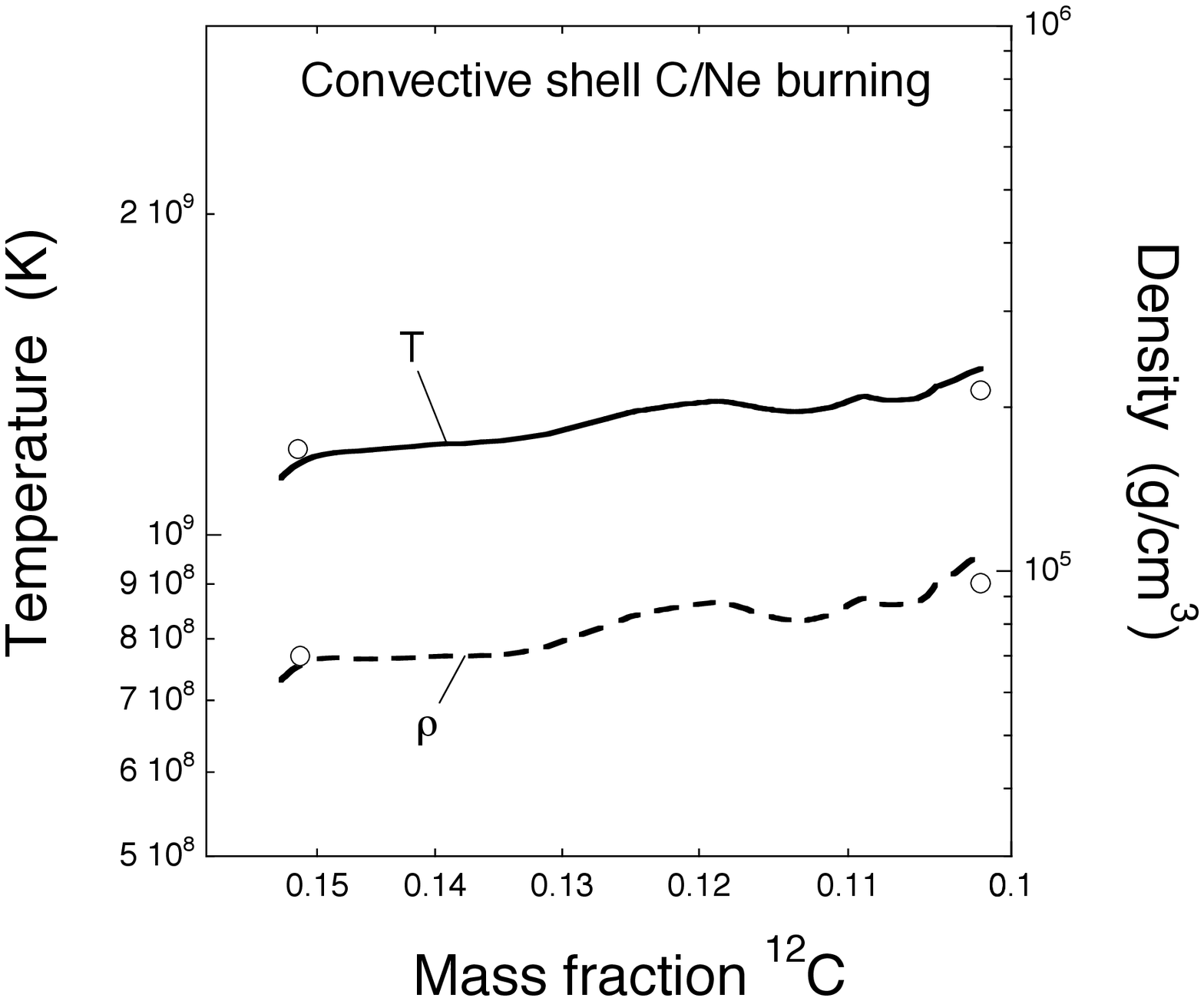}
\caption{Temperature-density evolution for convective shell C/Ne burning. The results were obtained from a model of a 60$M_\odot$ star (Limongi \& Chieffi 2006), but the time scale is shortened in the present work (see text). The profile approximates the evolution of the hottest and deepest zone of the C/Ne convective shell. For comparison, the circles indicate the temperature and density values that are directly obtained from the stellar evolution calculations. Time increases from left to right. \label{figconvtrho}
}
\end{figure}

Preliminary studies of the impact of nuclear uncertainties on pre-explosive $^{26}$Al yields can be found in Baldovin, Pignatari \& Gallino (2006). These authors performed post-processing studies using a schematic one-zone model consisting of two phases: a constant temperature of $T\approx1.1$ GK until the $^{12}$C mass fraction decreases from an inital value of 0.18 to 0.10 for phase 1, and a constant temperature of $T\approx1.3$ GK until the $^{12}$C mass fraction reaches a value of 0.050 for phase 2. In the present work we proceeded as follows. Initially, we extracted the temperature-density-time evolution of the deepest and hottest zone of the convective C/Ne burning shell from a stellar evolution model of a 60 $M_\odot$ star with initial solar metallicity (Limongi \& Chieffi 2006). This profile extended in time from the formation of the shell until a time of $t=3.15\times10^{6}$ s. Using this $T-\rho$ profile directly in a post-processing study would greatly distort the nucleosynthesis prediction, since the effects of convection are not taken properly into account. On the one hand, convection constantly carries fresh fuel (here $^{12}$C) into the burning region, while, on the other hand, it transports fragile nuclei from the burning region to cooler layers where they survive for a longer period of time. Therefore, in a stellar evolution calculation, convection has the effect of lengthening considerably the duration of nuclear burning. If we would use this profile directly in a post-processing simulation, then the initial $^{12}$C fuel, for example, would be destroyed much faster than in the actual stellar evolution calculation. After some trial attempts, we found that compressing the time axis of the original $T-\rho$ profile by a factor of 60 gives results that are consistent with the stellar evolution calculations. Clearly, this large scaling factor reflects the strong effects of convection during C/Ne shell burning. The results are shown in Fig. \ref{figconvtrho}, displaying the temperature and density dependence on the $^{12}$C mass fraction as solid and dashed lines, respectively. For comparison, the circles indicate the corresponding values from the stellar evolution calculations. The good agreement is encouraging and thus we used the scaled $T-\rho$ profile for our post-processing study. At the beginning of the burning, when $X_i(^{12}C)=0.15$, temperature and density start at values of $T=1.13$ GK and $\rho=6.3\times10^4$ g/cm$^3$, respectively. The profile extends over a time period of $t=5.24\times10^{4}$ s, when $X_f(^{12}C)=0.10$, and ends with values of $T=1.44$ GK and $\rho=1.1\times10^5$ g/cm$^3$. 

\begin{figure*}
\includegraphics[scale=1.0]{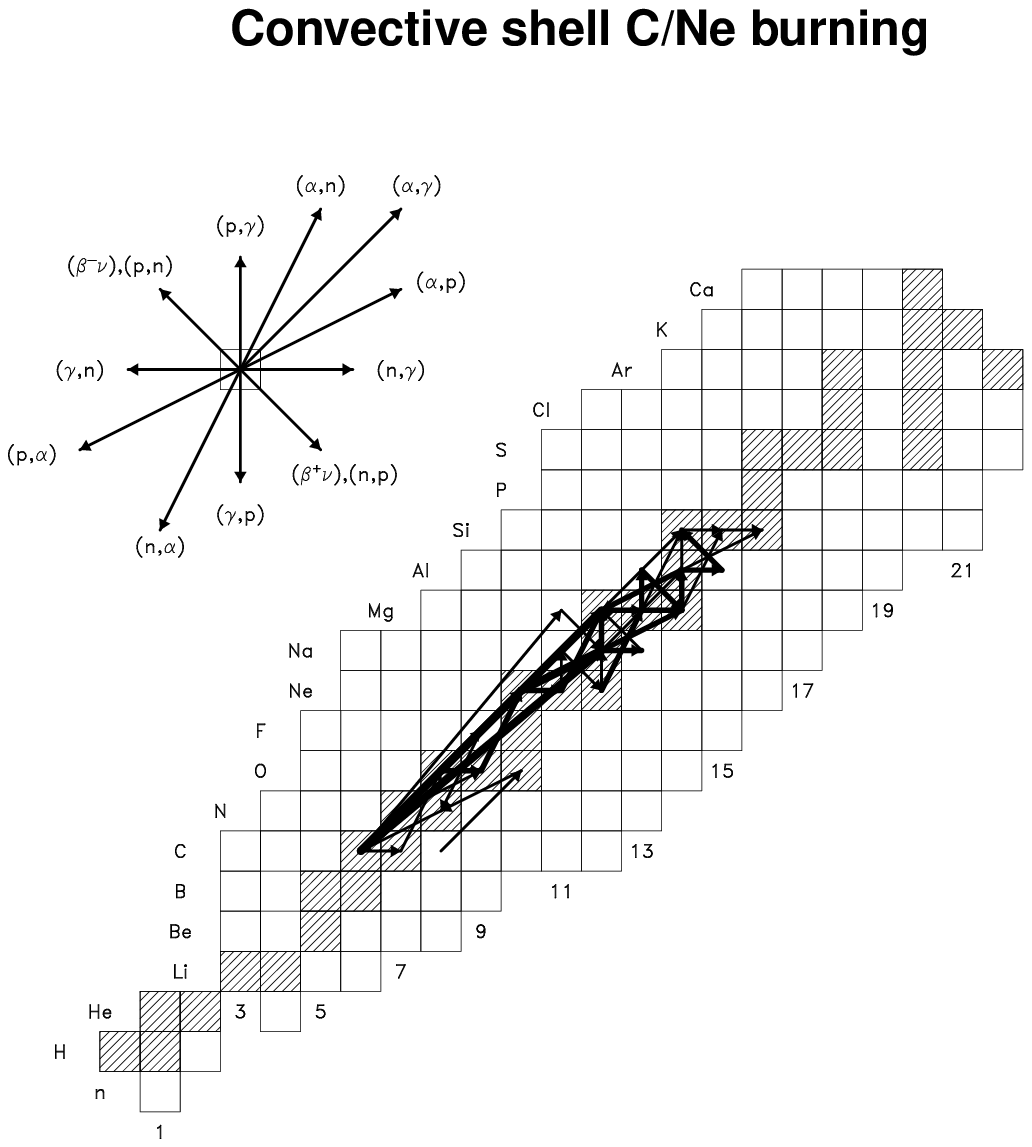}
\caption{Net abundance flows, obtained for a post-processing network calculation of convective shell C/Ne burning, integrated over a total running time of $t=5.2\times10^4$ s, when the $^{12}$C mass fraction has decreased to 0.097. The $T$-$\rho$ profile for this simulation is shown in Fig. \ref{figconvtrho}. The network consists of all nuclides shown as squares. The strongest net abundance flows, i.e., those within one, two, and three orders of magnitude of the maximum flow, are displayed by the thickest arrows, arrows of intermediate thickness, and the thinnest arrows, respectively. Thermal equilibrium for $^{26}$Al has been assumed (i.e., the network contains only a single species, $^{26}$Al$^t$).\label{figconvCflow}}
\end{figure*}

\begin{figure}
\includegraphics[scale=.49]{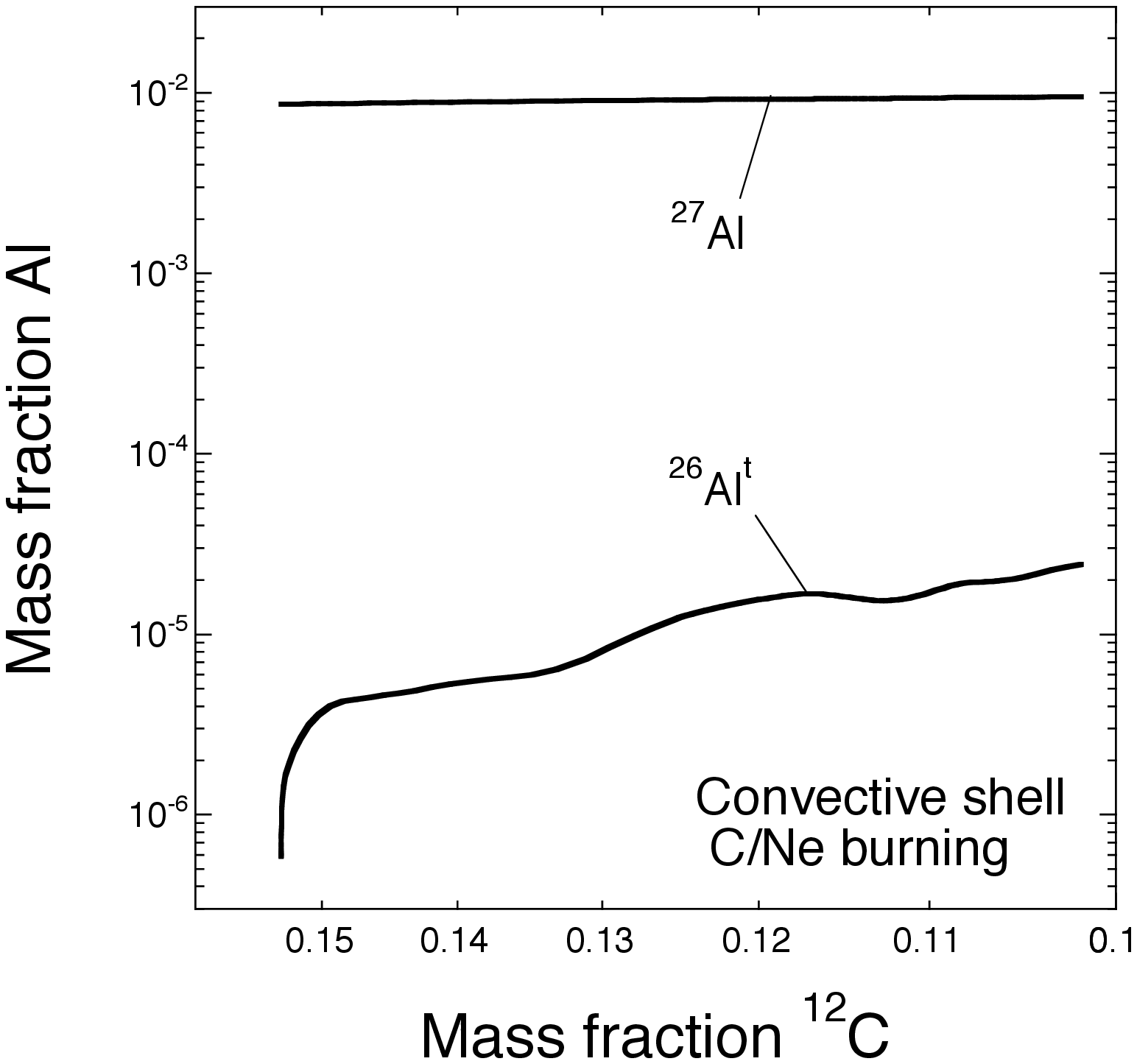}
\caption{Abundance evolution (by mass) of $^{26}$Al$^t$ and $^{27}$Al during convective shell C/Ne burning. The $T$-$\rho$ profile for this post-processing network simulation is shown in Fig. \ref{figconvtrho}. The calculation assumes a single species of (thermalized) $^{26}$Al. Time increases from left to right.
\label{figconvCalal}}
\end{figure}

At the beginning of the burning, the most abundant nuclides are (in order) $^{16}$O, $^{12}$C and $^{20}$Ne (Tab. \ref{tababun}). We assume at this stage thermal equilibrium for $^{26}$Al, i.e., the network contains only a single species, $^{26}$Al$^t$. The net abundance flows, integrated over a total running time of $t=5.24\times10^{4}$ s, for the standard calculation are displayed in Fig. \ref{figconvCflow}. The strongest net flows belong to the {\it primary} carbon burning reactions $^{12}$C($^{12}$C,$\alpha$)$^{20}$Ne and $^{12}$C($^{12}$C,p)$^{23}$Na, and to the {\it secondary} reactions $^{16}$O($\alpha$,$\gamma$)$^{20}$Ne and $^{23}$Na(p,$\alpha$)$^{20}$Ne that are initiated by the released light particles from the primary reactions. Near the end of convective shell C/Ne burning the mass fractions of protons, $\alpha$-particles and neutrons reach maximum values of 9.5$\times$10$^{-14}$, 5.0$\times$10$^{-9}$ and 2.9$\times$10$^{-16}$, respectively. The main direct process of $^{26}$Al$^t$ synthesis, in terms of the net abundance flow, is $^{25}$Mg(p,$\gamma$)$^{26}$Al$^t$, with $^{25}$Mg produced by the $^{22}$Ne($\alpha$,n)$^{25}$Mg and $^{24}$Mg(n,$\gamma$)$^{25}$Mg reactions. The main neutron source is $^{22}$Ne($\alpha$,n)$^{25}$Mg. On the other hand, $^{26}$Al$^t$ is mainly destroyed via the $\beta$-decay $^{26}$Al$^t\rightarrow^{26}$Mg (see column 2 of Tab. \ref{tblaldecay}). In particular, the neutron abundance is too low in the standard calculation for the destruction reactions $^{26}$Al$^t$(n,p)$^{26}$Mg and $^{26}$Al$^t$(n,$\alpha$)$^{23}$Na to compete successfully with the $\beta$-decay of $^{26}$Al$^t$. The abundance evolutions of the species $^{26}$Al$^t$ and $^{27}$Al are shown in Fig. \ref{figconvCalal}. While the abundance of the latter nuclide is approximately constant throughout the calculation, the abundance of the former species increases by more than an order of magnitude over the course of the burning. The abundance ratio, $X(^{26}Al^t)/X(^{27}Al)$, increases from an initial value of $6.7\times10^{-5}$ to a final value of $2.5\times10^{-3}$ (see Tab. \ref{tababun}).

\subsubsection{Reaction rate variations}
Subsequently, the rates of 66 pairs of forward and reverse reactions were varied. Those reactions whose rate changes have the strongest effect on the final $^{26}$Al yield (i.e., at the end of the calculation, when $X_f(^{12}C)=0.10$) are listed in Tab. \ref{tabcC}. All other rate changes, as well as those labeled by ``..." in the table, produced $^{26}$Al$^t$ abundance changes of less than 20\%. The reactions are listed in approximate order of importance, as measured by their impact on the final $^{26}$Al abundance. The last two columns display the source of the rate and the reported rate uncertainty at a temperature of $\approx$1.4 GK near the end of the calculation. Disregarding at first the actual rate uncertainties, the four reactions with the strongest impact on $^{26}$Al$^t$ nucleosynthesis are: $^{25}$Mg(p,$\gamma$)$^{26}$Al$^t$, $^{26}$Al$^t$(n,p)$^{26}$Mg, $^{23}$Na(p,$\alpha$)$^{20}$Ne and $^{23}$Na($\alpha$,p)$^{26}$Mg. The first reaction produces $^{26}$Al$^t$, while multiplying the rate of the second reaction by a factor of 10 would make it the dominant $^{26}$Al$^t$ destruction process, at the cost of the $\beta$-decay of $^{26}$Al$^t$. The third reaction is the main consumer of free protons. When the rate of the $^{23}$Na(p,$\alpha$)$^{20}$Ne reaction is increased, the number of available protons decreases and, consequently, fewer $^{25}$Mg nuclei can be converted to $^{26}$Al$^t$. The fourth reaction represents the second most important proton-generating process (after the primary $^{12}$C($^{12}$C,p)$^{23}$Na reaction). When the $^{23}$Na($\alpha$,p)$^{26}$Mg reaction rate is increased, more protons are available for producing $^{26}$Al$^t$ from $^{25}$Mg. 

\begin{deluxetable}{llllllc}
\tablewidth{0pt}
\tablecaption{Factor changes of final $^{26}$Al$^t$ abundance resulting from reaction rate variations for convective shell C/Ne burning\tablenotemark{a}, assuming thermal equilibrium for $^{26}$Al\label{tabcC}}
\tablehead{
\colhead{Reaction\tablenotemark{b}} &  \multicolumn{4}{c}{Rate multiplied by}  \nl 
\cline{2-5}
&  \colhead{10} & \colhead{2} & \colhead{0.5} & \colhead{0.1}   & \colhead{Source\tablenotemark{c}}    &  \colhead{Uncertainty\tablenotemark{d}}   
}
\startdata
$^{23}$Na(p,$\alpha$)$^{20}$Ne		        &  0.15        &  0.61          &  1.6      & 4.2                    &  il10     &  6\% \\
$^{26}$Al$^t$(n,p)$^{26}$Mg			        &  0.16        &  0.65          &  1.4      &  1.9                   &  present   &   \\
$^{25}$Mg(p,$\gamma$)$^{26}$Al$^t$           &  6.2          &  2.0            &  0.46      & 0.10                 &  il10     &  5\% \\
$^{23}$Na($\alpha$,p)$^{26}$Mg		        & 3.0          &  1.3             &  \nodata       &  0.71         &  rath     &   \\
$^{26}$Mg($\alpha$,n)$^{29}$Si		        &  0.40       &   0.83          &  \nodata      &  1.3            &  nacr    &  29\% \\
$^{24}$Mg(n,$\gamma$)$^{25}$Mg		        &  2.1         &  1.3              & \nodata       &  0.70         &  ka02   &   \\
$^{26}$Al$^t$(n,$\alpha$)$^{23}$Na	        &  0.54       &  0.79            &  \nodata      &  \nodata    &  present   &   \\
$^{16}$O($\alpha$,$\gamma$)$^{20}$Ne       & \nodata    & 0.83            & 1.3               &  1.7           &  il10     &  14\% \\
$^{25}$Mg($\alpha$,n)$^{28}$Si		        &  0.42       &  \nodata      & \nodata       &  \nodata     &  nacr    &  59\% \\
$^{12}$C($^{12}$C,n)$^{23}$Mg		        &  0.46       &  \nodata      &  \nodata      &  \nodata     &  da77   &   \\
$^{27}$Al(n,$\gamma$)$^{28}$Al                    &  1.7         &  \nodata       &  \nodata      &   \nodata    &  ka02   &   \\
$^{25}$Mg(n,$\gamma$)$^{26}$Mg                &  1.3         &  \nodata       &  \nodata      &   \nodata    &  ka02   &   \\
$^{26}$Mg(p,$\gamma$)$^{27}$Al                  &  0.71       &  \nodata       & \nodata       &  \nodata     &  il10     &  5\% \\
$^{27}$Al(p,$\alpha$)$^{24}$Mg                     &  0.79       &  \nodata       &  \nodata      &  \nodata     &  il10     &  7\% \\
\enddata
\tablenotetext{a}{The temperature-density-time profile is extracted from a stellar evolution calculation of a 60$M_\odot$ star with initial solar metallicity, see Limongi \& Chieffi (2006).}
\tablenotetext{b}{In total, the rates of 66 different reactions were varied. Listed are only those reactions whose rate changes have the strongest effect on the $^{26}$Al$^t$ yield. All other rate changes, as well as those labeled by ``...", produced abundance changes of less than 20\%. The reactions are listed in approximate order of importance. Thermal equilibrium for $^{26}$Al has been assumed, i.e., the network contains only a single species, $^{26}$Al$^t$.}
\tablenotetext{c}{Reaction rate references: (nacr) Angulo et al. 1999 (NACRE); (ka02) Dillmann et al. (2006) (KADoNiS v0.2); (rath) Rauscher \& Thielemann (2000); (il10) Iliadis et al. (2010); (present) hybrid rate, see Appendix \ref{al26na} and \ref{al26np}; (da77) total $^{12}$C+$^{12}$C rate from Caughlan \& Fowler (1988), with neutron branching ratio adopted from Dayras et al. (1977), see Appendix \ref{c12c12n}.}
\tablenotetext{d}{Reaction rate uncertainty near a temperature of 1.4 GK, at the end of the calculation; no entry implies that the rate uncertainty is difficult to quantify (see text).}
\end{deluxetable}

Other important rate variations that impact the final $^{26}$Al$^t$ abundance arise from the reactions $^{25}$Mg($\alpha$,n)$^{28}$Si, $^{24}$Mg(n,$\gamma$)$^{25}$Mg and $^{26}$Al$^t$(n,$\alpha$)$^{23}$Na, which have already been discussed in $\S$ \ref{xNevary}. In addition, the reactions $^{16}$O($\alpha$,$\gamma$)$^{20}$Ne, $^{12}$C($^{12}$C,n)$^{23}$Mg and $^{26}$Mg($\alpha$,n)$^{29}$Si play an important role. Decreasing the rate of the first reaction will consume fewer $\alpha$-particles, thus increasing the production of neutrons (for converting $^{24}$Mg to $^{25}$Mg seed) via $^{22}$Ne($\alpha$,n)$^{25}$Mg. Increasing the rate of the second reaction produces an additional burst of neutrons towards the very end of the burning, thereby destroying more $^{26}$Al$^t$ nuclei via the (n,p) and (n,$\alpha$) reactions. For specific comments on the $^{12}$C($^{12}$C,n)$^{23}$Mg reaction, see Appendix \ref{c12c12n}. Similar arguments apply to the $^{26}$Mg($\alpha$,n)$^{29}$Si reaction. There are four more reactions listed in Tab. \ref{tabcC}, $^{27}$Al(n,$\gamma$)$^{28}$Al, $^{25}$Mg(n,$\gamma$)$^{26}$Mg, $^{26}$Mg(p,$\gamma$)$^{27}$Al and $^{27}$Al(p,$\alpha$)$^{24}$Mg, which impact the final $^{26}$Al$^t$ abundance. The only reactions that we found to influence the $^{26}$Al$^t$ yield but are not listed in Tab. \ref{tabcC} are $^{12}$C($^{12}$C,p)$^{23}$Na and $^{12}$C($^{12}$C,$\alpha$)$^{20}$Ne. Varying these rates by a factor of 10 changes the $^{26}$Al$^t$ abundance by a factor of $\approx$2. However, these reactions drive carbon burning and thus strongly influence the nuclear energy generation. Therefore, varying this rate in a post-processing study is not very meaningful. Nevertheless, the effect appears to be relatively small.

\subsubsection{Reaction rate uncertainties\label{uncertConvCb}}
Of the 14 reactions listed in Tab. \ref{tabcC}, the rates of: (i) 5 reactions are available from the Monte Carlo procedure (Iliadis et al. 2010; $\S$ \ref{physlib}); (ii) 3 reactions are adopted from the statistical model (Rauscher \& Thielemann 2000); (iii) 3 reactions are obtained from KADoNiS v0.2 (Dillmann et al. 2006); and (iv) 2 reactions are adopted from NACRE (Angulo et al. 1999). Only the rates of the $^{12}$C($^{12}$C,n)$^{23}$Mg reaction are partially based (i.e., the {\it total} $^{12}$C+$^{12}$C rate) on the information provided in Caughlan \& Fowler (1988), see Appendix \ref{c12c12n}.

Reaction rate uncertainties are listed for a temperature of $T=1.4$ GK, near the end of the burning, in the last column of Tab. \ref{tabcC}, when reported in the original source. The uncertainties for $^{23}$Na(p,$\alpha$)$^{20}$Ne, $^{25}$Mg(p,$\gamma$)$^{26}$Al$^t$, $^{16}$O($\alpha$,$\gamma$)$^{20}$Ne, $^{26}$Mg(p,$\gamma$)$^{27}$Al and $^{27}$Al(p,$\alpha$)$^{24}$Mg are relatively small and, therefore, the rate estimates for these reactions seem sufficiently reliable at present. The $^{26}$Mg($\alpha$,n)$^{29}$Si and $^{25}$Mg($\alpha$,n)$^{28}$Si reactions are listed with rather large rate uncertainties (29\% and 59\%, respectively, according to Angulo et al. 1999) and, therefore, should be addressed in future work (see also Appendix \ref{mg25anapp}). No rate uncertainties are given for any of the other reactions listed in the table. These rates are derived, for example, from Hauser-Feshbach theory (Rauscher \& Thielemann 2000) or from the KADoNiS v0.2 evaluation (Dillmann et al. 2006) and uncertainties are difficult to quantify, as has already been discussed in $\S$ \ref{uncert}. Clearly, more reliable experimental rates for these reactions are urgently needed. Specific comments on the reactions $^{26}$Al$^t$(n,p)$^{26}$Mg, $^{26}$Al$^t$(n,$\alpha$)$^{23}$Na, $^{23}$Na($\alpha$,p)$^{26}$Mg and $^{12}$C($^{12}$C,n)$^{23}$Mg can be found in Appendix \ref{specrates}. 

\subsubsection{Thermal equilibration}
So far we assumed a single species of $^{26}$Al, implying thermal equilibrium ($^{26}$Al$^t$). We will now follow the equilibration numerically in the network calculation. Five different species of $^{26}$Al are incorporated into the network, as explained in $\S$ \ref{thermequi}. The required $\gamma$- and $\beta$-decay transitions between and from these levels are discussed in detail in Appendix \ref{appdec}. The additional reaction rates, involving $^{26}$Al$^g$ and $^{26}$Al$^m$ separately, are discussed in Appendix \ref{appreac}.

\begin{figure}
\includegraphics[scale=.55]{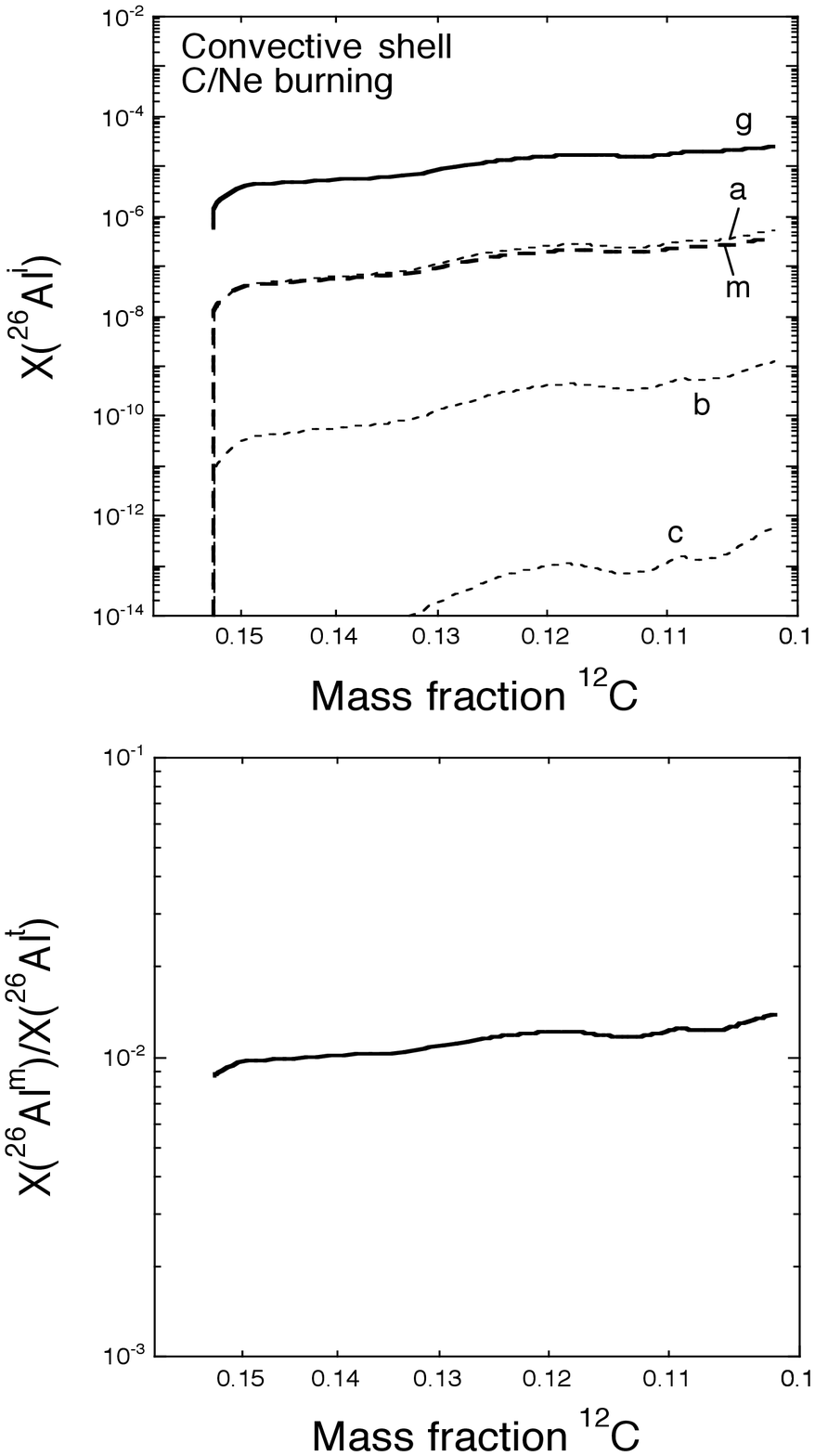}
\caption{Abundance evolution (by mass) of $^{26}$Al during convective shell C/Ne burning. The $T$-$\rho$ profile for this post-processing network simulation is shown in Fig. \ref{figconvtrho}. The calculation assumes five species of $^{26}$Al: ground state ($g$), isomeric state ($m$), and three excited levels ($a$, $b$, $c$); see Fig. \ref{figlevel}. The communication of the different $^{26}$Al levels via $\gamma$-ray transitions is explicitly taken into account. (Top) Abundance evolution of different $^{26}$Al species; (Bottom) Numerically simulated fraction of total $^{26}$Al abundance that resides in the isomeric state. The curve is indistinguishable from the one calculated assuming a Boltzmann distribution (i.e., thermal equilibrium).
\label{figconvCab}}
\end{figure}

As a first step, a standard post-processing network calculation (with recommended rates) is performed using the same temperature-density-time evolution as before (Fig. \ref{figxNetrho}). The final $^{26}$Al$^{g}$ abundance, when the $^{12}$C mass fraction has fallen to a value of $X_f(^{12}C)=0.10$, is found to be identical to our earlier result obtained assuming thermal equilibrium (Tab. \ref{tababun}). Thus the latter assumption seems to be justified. An impression can be gained from Fig. \ref{figconvCab}, showing the abundance evolution of $^{26}$Al levels. The top part displays the abundances for individual $^{26}$Al species and it is apparent that at any given time the $^{26}$Al$^g$ abundance dominates over those of the other species. The bottom part displays the fraction of the total $^{26}$Al abundance that resides in the isomeric state. This curve is directly obtained from the network calculation, but is indistinguishable from the one calculated assuming a Boltzmann distribution (i.e., thermal equilibrium).

Subsequently, the rates of 20 different interactions, together with their inverse processes, were varied individually by factors of 100, 10, 2, 0.5, 0.1 and 0.01. This list contained all nuclear reactions that produced and destroyed $^{26}$Al$^g$ or $^{26}$Al$^m$. It also included those $\beta$- and $\gamma$-ray decay rates of $^{26}$Al$^x$ levels that were estimated using the shell model, as described in $\S$ \ref{thermequi} and shown in Fig. \ref{figlevel}. Experimentally obtained $\beta$- and $\gamma$-ray decay rates have not been varied since their uncertainties are very small. The final $^{26}$Al$^g$ abundance, after each rate variation, was then compared to the standard calculation. The results are listed in Tab. \ref{tabcC5s}. It can be seen that of the 20 interactions only five, all of them reactions, influence the final $^{26}$Al$^g$ yield. In other words, even a variation by a factor of 100 of the shell-model based $\beta$- and $\gamma$-ray decay rates seems to have no effect on the $^{26}$Al$^g$ abundance.

\begin{deluxetable}{llllllllc}
\tablewidth{0pt}
\tablecaption{Factor changes of final $^{26}$Al$^g$ abundance resulting from reaction rate variations for convective shell C/Ne burning\tablenotemark{a}, assuming five species of $^{26}$Al\label{tabcC5s}}
\tablehead{
\colhead{Reaction\tablenotemark{b}} &  \multicolumn{6}{c}{Rate multiplied by}  \nl 
\cline{2-7}
&  \colhead{100} & \colhead{10} & \colhead{2} & \colhead{0.5} & \colhead{0.1} & \colhead{0.01}  & \colhead{Source\tablenotemark{c}}    &  \colhead{Uncertainty\tablenotemark{d}}   
}
\startdata
$^{26}$Al$^g$(n,p)$^{26}$Mg                    &  0.017  & 0.16       &  0.63         & 1.3         & 1.9          & 2.0          &  present   &    \\
$^{25}$Mg(p,$\gamma$)$^{26}$Al$^g$     &  2.9      & 5.4         &  1.5           & 0.63       & 0.35        & 0.29         &  il10     &   5\% \\
$^{25}$Mg(p,$\gamma$)$^{26}$Al$^m$    &  6.7      & 3.0         &  \nodata    & \nodata  & 0.75        & 0.71        &  il10      &   6\% \\
$^{26}$Al$^g$(n,$\alpha$)$^{23}$Na         &  0.12     & 0.54      &  \nodata    & \nodata  &  \nodata  &  \nodata   &  present   &   \\
$^{26}$Al$^m$(n,p)$^{26}$Mg                   &  0.58     & \nodata  &  \nodata    & \nodata  &  \nodata  &  \nodata   &  present   &    \\
\enddata
\tablenotetext{a}{The temperature-density-time profile is extracted from a stellar evolution calculation of a 60$M_\odot$ star with initial solar metallicity, see Limongi \& Chieffi (2006).}
\tablenotetext{b}{In total, the rates of 20 different reactions producing or destroying $^{26}$Al$^g$ and $^{26}$Al$^m$ were varied. Listed are only those reactions whose rate changes have the strongest effect on the $^{26}$Al$^g$ yield. All other rate changes, as well as those labeled by ``...", produced abundance changes of less than 20\%. The reactions are listed in approximate order of importance. No thermal equilibrium for $^{26}$Al has been explicitly assumed, i.e., the network contains five different species ($^{26}$Al$^g$, $^{26}$Al$^m$, $^{26}$Al$^a$, $^{26}$Al$^b$, $^{26}$Al$^c$) and takes the interactions between them into account.}
\tablenotetext{c}{Reaction rate references: (il10) Iliadis et al. 2010; (present) hybrid rate, see Appendix \ref{al26na} and \ref{al26np}. In the latter two cases, we assumed that the rate involving $^{26}$Al$^g$ or $^{26}$Al$^m$ is the same as the rate for $^{26}$Al$^t$ (see comments in Appendix \ref{appreac}).}
\tablenotetext{d}{Reaction rate uncertainty near a temperature of 1.4 GK, at the end of the calculation; no entry implies that the rate uncertainty is difficult to quantify.}
\end{deluxetable}

The five reactions displayed in Tab. \ref{tabcC5s} are listed in approximate order of importance. For example, increasing the $^{25}$Mg(p,$\gamma$)$^{26}$Al$^g$ rate by a factor of 2 will enhance the final $^{26}$Al yield by $\approx$50\%. However, this rate is based on experimental information (Iliadis et al. 2010) and their Monte Carlo uncertainty is predicted to amount to only 5\% near $T=1.4$ GK. As was the case for explosive Ne/C burning ($\S$ \ref{secthermxNe}), we found that even a factor of 100 variation in the rates of these five reactions has no impact on the thermal equilibrium abundance ratio of $^{26}$Al$^m$ and $^{26}$Al$^g$ (Fig. \ref{figconvCab}). Comparison of the factor changes listed in Tab. \ref{tabcC5s} with those of Tab. \ref{tabcC} reveals that these five reactions impact the final $^{26}$Al abundance by similar amounts, no matter if a single (thermalized) or five species of $^{26}$Al are used in the simulation. In conclusion, $^{26}$Al is in thermal equilibrium during convective shell C/Ne burning and, therefore, there is no need to introduce the extra complication of five $^{26}$Al levels, and their mutual interactions, into the reaction network.

\subsection{Convective core H burning}
\subsubsection{Standard calculation}
\begin{figure}
\includegraphics[scale=0.40]{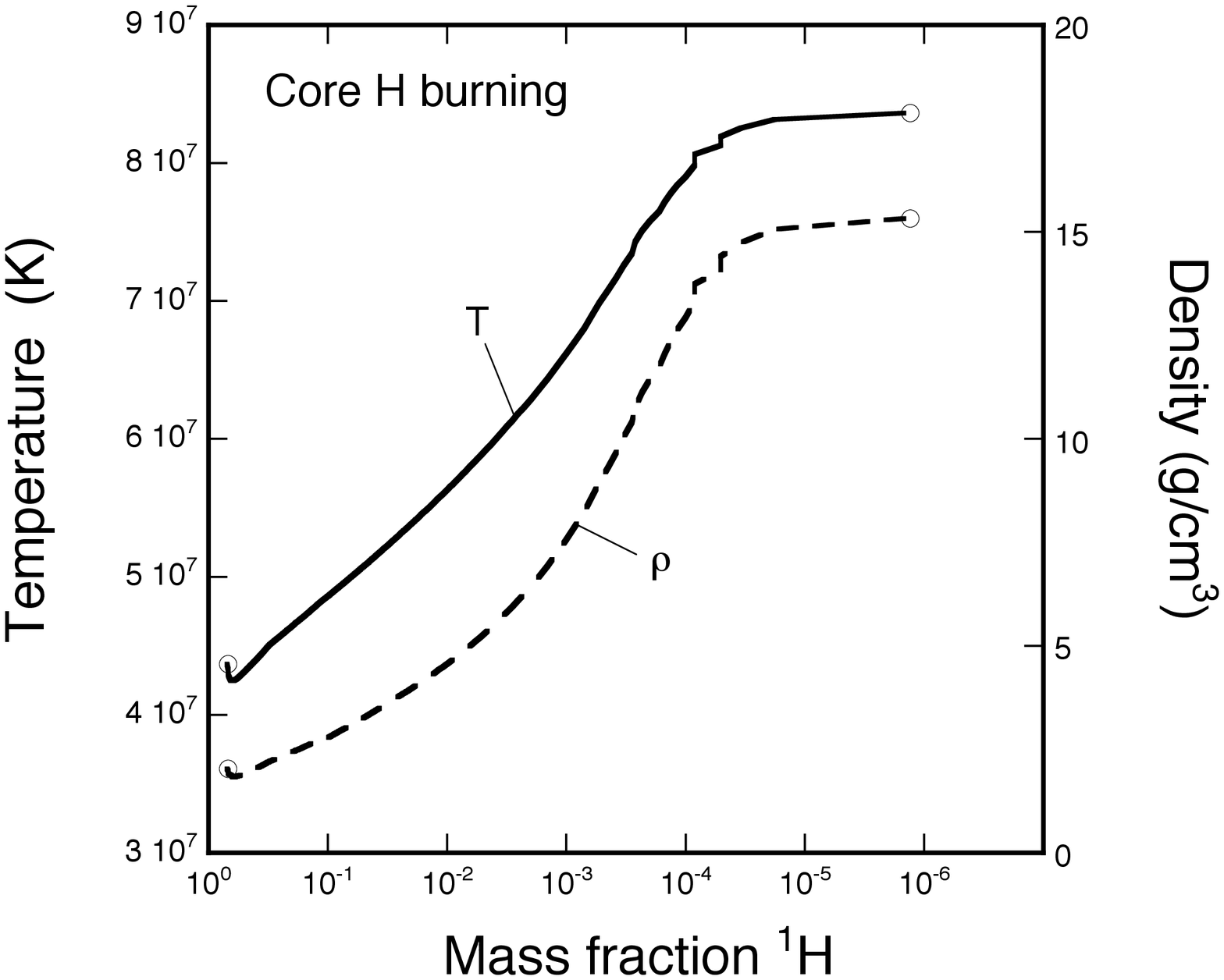}
\caption{Temperature-density evolution for convective core H burning. The results were obtained from a model of a 80M$_{\odot}$ star (Limongi \& Chieffi 2006), but the time scale is shortened in the present work (see text). The circles indicate the temperature and density values that are directly obtained from the above stellar evolution calculations. Time increases from left to right.}
\label{fig:coreHprofile}
\end{figure}

We have performed post-processing calculations using the temperature-density profile for convective core H-burning in an 80 M$_{\odot}$ star of solar initial composition (Limongi \& Chieffi 2006). Following the procedure adopted for convective shell C/Ne burning, we have artificially shortened the burning time
in our calculation so that the time evolution of the hydrogen fuel would closely follow that for the stellar evolution calculation. For this profile, compressing the time axis by a factor of 17 gave consistent results, as shown in Fig. \ref{fig:coreHprofile}. Burning starts at $X_i$($^1H$) = 0.70, $T$ =  0.044 GK and $\rho$ = 2.03 g/cm$^{3}$. The modified profile extends for $t=5.85 \times 10^{12}$ s, at which time $X_f$($^1H$) = 1.4 $\times$ 10$^{-6}$, $T$ =  0.088 GK and $\rho$ = 17.9 g/cm$^{3}$. Note, that in this case $T$ and $\rho$ refer to the values at the center of the star.

\begin{figure*}
\includegraphics[scale=1.0]{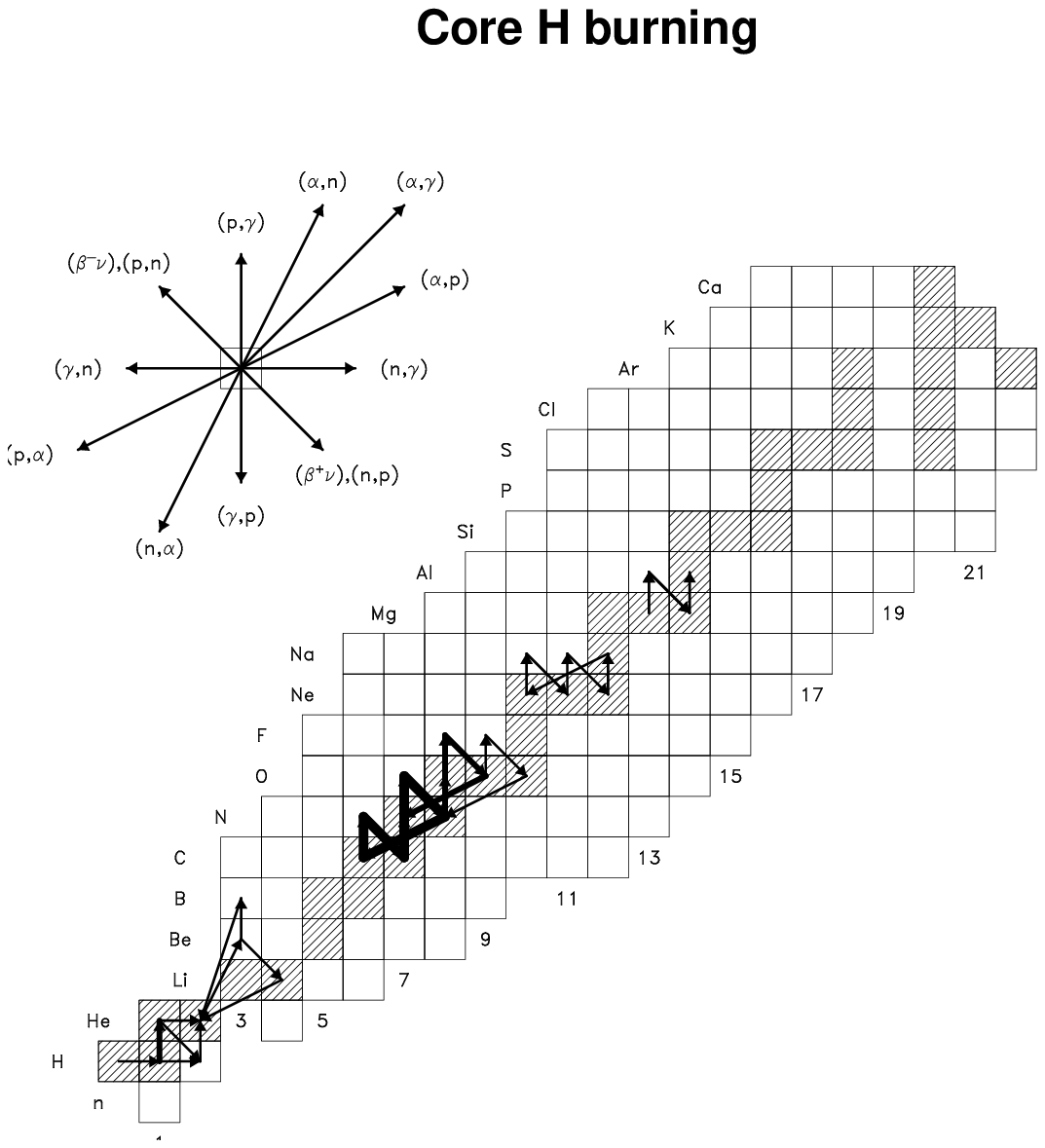}
\caption{Net abundance flows, obtained for a post-processing network calculation of convective core H burning, integrated over a total running time of t = $5.9\times10^{12}$ s, when the $^{1}$H mass fraction has decreased to 1.3 $\times$ 10$^{-6}$. The T-$\rho$ profile for this simulation is shown in Fig. \ref{fig:coreHprofile}. The network consists of all nuclides shown as squares.
The strongest net abundance flows, i.e., those within two, four, and six orders of magnitude of the maximum flow, are displayed by the thickest arrows, arrows of intermediate thickness, and the thinnest arrows, respectively.\label{fig:coreHflow}}
\end{figure*}

\begin{figure}
\includegraphics[scale=0.44]{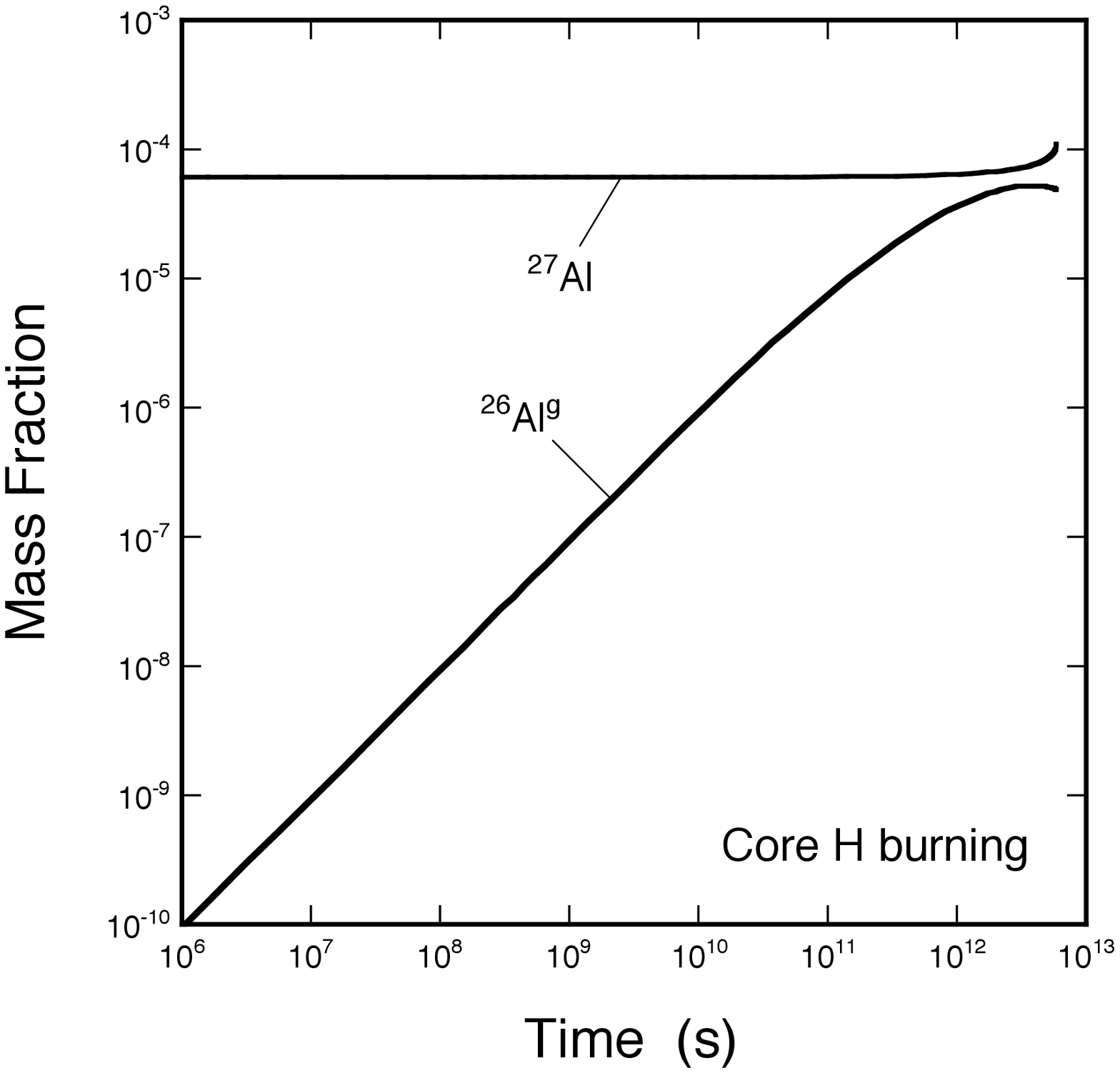}
\caption{Abundance evolution (by mass) of $^{26}$Al$^{g}$ and $^{27}$Al during convective core H burning. The T-$\rho$ profile for this post-processing simulation is shown in Fig. \ref{fig:coreHprofile}.}
\label{fig:coreHabund}
\end{figure}

Our standard calculation assumes that $^{26}$Al$^{g}$ and $^{26}$Al$^{m}$ are distinct species. The net abundance flows are shown in Fig. \ref{fig:coreHflow} and not surprisingly, the strongest are within the CNO cycles. Both the ground state of $^{26}$Al and the isomeric level are produced via the $^{25}$Mg(p,$\gamma$)$^{26}$Al reaction. For most of the burning period, $^{26}$Al is produced from the reservoir of initial $^{25}$Mg and it is not until the very late stages of burning that $^{25}$Mg is replenished through $^{24}$Mg(p,$\gamma$)$^{25}$Al and the subsequent $\beta$-decay of $^{25}$Al. The primary destruction route for the isomer is $\beta$-decay to $^{26}$Mg, whereas the ground state is destroyed via $^{26}$Al$^{g}$(p,$\gamma$)$^{27}$Si. The abundance evolution of $^{26}$Al$^{g}$ and $^{27}$Al is shown in Fig. \ref{fig:coreHabund}. The abundance of $^{27}$Al is essentially constant at X($^{27}$Al) = 6.1 $\times$ 10$^{-5}$ until late times while $^{26}$Al$^{g}$ grows to a maximum of X($^{26}$Al$^{g}$) = 5.2 $\times$ 10$^{-5}$ before dropping to X($^{26}$Al$^{g}$) = 4.9 $\times$ 10$^{-5}$ at the end of burning (see Tab. 1). The ratio $^{26}$Al$^{g}$/$^{27}$Al shows a similar behavior, reaching a maximum of 0.73 with a final value of 0.45.
 

\subsubsection{Reaction rate variations, thermal equilibration and uncertainties}
The rates of 26 pairs of forward and reverse reactions were varied and those reactions whose rate changes have the strongest effect on the final abundance of $^{26}$Al$^{g}$ (i.e., at the end of the calculation, when X$_{H}$ = 1.3 $\times$ 10$^{-6}$) are listed in Tab. \ref{tab:coreHrates}. All other rate changes, as well as those labeled by ``..." in the table, changed the $^{26}$Al$^{g}$ abundance by less than 20\%. The reactions are listed in approximate order of importance, as measured by their impact on the final $^{26}$Al$^{g}$  abundance. The last two columns display the source of the rate and the reported rate uncertainty at a temperature of $\approx$ 0.09 GK, near the end of the calculation. For these calculations, $^{26}$Al$^{g}$ and $^{26}$Al$^{m}$ are considered to be separate species. It is not surprising that the $^{25}$Mg(p,$\gamma$)$^{26}$Al$^{g}$ reaction has the largest impact on the final $^{26}$Al$^{g}$  abundance, but it is interesting that variations in the rate of $^{25}$Mg(p,$\gamma$)$^{26}$Al$^{m}$ also affect the $^{26}$Al$^{g}$  abundance. This is because these two reactions are the only significant destruction mechanisms for $^{25}$Mg and thus $^{25}$Mg(p,$\gamma$)$^{26}$Al$^{m}$ lowers the amount of $^{25}$Mg available to be converted to $^{26}$Al$^{g}$. The third reaction, $^{26}$Al$^{g}$(p,$\gamma$)$^{27}$Si, is the major destruction route for $^{26}$Al$^{g}$. Finally, the  $^{16}$O(p,$\gamma$)$^{17}$F reaction affects $^{26}$Al production by reducing the abundance of free protons. At no point during the calculation did the temperature reach a point where $^{26}$Al$^{g}$ and $^{26}$Al$^{m}$ could communicate through thermal excitations. This was verified by including the three mediating levels discussed in $\S$ 2.3 in a second series of network calculations and no change was seen in the final abundance of $^{26}$Al$^{g}$.

The rates for all of the reactions listed in Tab. \ref{tab:coreHrates} were obtained using the Monte Carlo procedure (Iliadis et al. 2010; $\S$ 2.2) and for the temperatures encountered in convective core H-burning, all of these are based on experimental data. The uncertainties quoted are at the 1-$\sigma$ level for lognormal probability density functions. Judging from the entries in Tab. \ref{tab:coreHrates}, none of these reactions will impact the abundance of $^{26}$Al$^{g}$ if their rates are varied within a factor of 2 from their recommended values. Given the quoted uncertainties, a factor of 2 corresponds to confidence intervals of 97.9\% for $^{25}$Mg(p,$\gamma$)$^{26}$Al$^{g}$, 98.4\% for $^{25}$Mg(p,$\gamma$)$^{26}$Al$^{m}$, 99.3\% for $^{26}$Al$^{g}$(p,$\gamma$)$^{27}$Si and $\approx$ 100\% for  $^{16}$O(p,$\gamma$)$^{17}$F. In other words it is unlikely that the rates for any of these reactions will be a factor of 2 away from the recommended values. This points to the utility of uncertainties with statistical significance.  Therefore, we conclude that the rates for the reactions that determine the abundance of $^{26}$Al$^{g}$ during convective core H-burning are known with sufficient precision. 

\begin{deluxetable}{llllllllc}
\tablewidth{0pt}
\tablecaption{Factor changes of final $^{26}$Al$^{g}$ abundance resulting from reaction rate variations for convective core H burning\tablenotemark{a}, assuming two species of $^{26}$Al\label{tab:coreHrates}}
\tablehead{
\colhead{Reaction\tablenotemark{b}} &  \multicolumn{6}{c}{Rate multiplied by}  \nl 
\cline{2-7}
&  \colhead{100} & \colhead{10} & \colhead{2} & \colhead{0.5} & \colhead{0.1} & \colhead{0.01}  & \colhead{Source\tablenotemark{c}}    &  \colhead{Uncertainty\tablenotemark{d}}   
}
\startdata
$^{26}$Al$^{g}$(p,$\gamma$)$^{27}$Si & 0.0035 & 0.55 & \nodata & \nodata & \nodata & \nodata & Il10 & 31\% \\
$^{25}$Mg(p,$\gamma$)$^{26}$Al$^{g}$ & 1.20 & 1.20 & \nodata & \nodata & 0.33 & 0.039 & Il10& 35\% \\
$^{25}$Mg(p,$\gamma$)$^{26}$Al$^{m}$ & 0.049 & 0.37 & \nodata & \nodata & 1.20 & 1.20 & Il10 & 35\% \\
$^{16}$O(p,$\gamma$)$^{17}$F & \nodata & \nodata & \nodata & \nodata & \nodata & 1.70 & Il10 & 7\%\\
\enddata
\tablenotetext{a}{The temperature-density-time profile is extracted from a stellar evolution calculation of a 80M$_{\odot}$ star with initial solar metallicity, see Limongi \& Chieffi (2006).}
\tablenotetext{b}{In total, the rates of 26 different reactions were varied. Listed are
only those reactions whose rate changes have the strongest effect on the
$^{26}$Al$^{g}$ yield. All other rate changes, as well as those labeled by ``...",
produced abundance changes of less than 20\%. The reactions are listed
in approximate order of importance.}
\tablenotetext{c}{Reaction rate reference: (il10) Iliadis et al. (2010).}
\tablenotetext{d}{Reaction rate uncertainty near a temperature of 0.09 GK, at the end of the calculation.}
\end{deluxetable}

\section{Summary}
We presented a comprehensive investigation of the impact of nuclear reaction rate uncertainties on $^{26}$Al production in massive stars. In such stars, $^{26}$Al is likely produced in three distinct sites: (i) during core collapse via explosive Ne/C burning; (ii) during pre-supernova stages in the C/Ne convective shell, where a fraction of the $^{26}$Al survives the subsequent explosion and is ejected into the interstellar medium; and (iii) in Wolf-Rayet stars, that experience such a strong mass loss that even layers located within the H convective core, hence significantly enriched in $^{26}$Al, are ejected into the interstellar medium. These $^{26}$Al production mechanisms were recently analyzed in detail by Limongi \& Chieffi (2006). From their stellar evolution models, we extracted representative temperature-density-time profiles and executed a large number of post-processing reaction network sensitivity calculations. The general strategy consisted of varying the rates of many reactions individually by different factors (in this work, 10, 2, 0.5 and 0.1) and to analyze the impact of each individual reaction rate change on the final $^{26}$Al yields. Our results are important for quantifying the influence of current reaction rate uncertainties on predicted $^{26}$Al yields, and for the motivation of future laboratory measurements.

There are a number of novel aspects about the present work. First, we employed a new-generation library of nuclear reaction and weak interaction rates, called STARLIB. This library contains a recent evaluation of experimental Monte Carlo reaction rates (Iliadis et al. 2010). Besides recommended reaction rates for a grid of temperature values between 1 MK and 10 GK, the library includes in addition for many reactions the rate uncertainty factor at each temperature. This work represents the first application of STARLIB. Second, we carefully investigate the equilibration effects of $^{26}$Al. At least two species of $^{26}$Al take part in the nucleosynthesis, the ground state and the isomeric state. In all previous massive star investigations, either a single species or two species of $^{26}$Al were taken into account, depending on whether thermal equilibrium is achieved or not. These are two extreme assumptions and in a hot stellar plasma the ground and isomeric state may ``communicate" via $\gamma$-ray transitions involving higher-lying $^{26}$Al levels.

Some of our results are summarized in Tab. \ref{tabsum}, listing those nuclear reactions that significantly impact $^{26}$Al synthesis in massive stars. The reactions are listed in approximate order of importance. The reader should consult Tabs. \ref{tabxNe}, \ref{tabcC} and \ref{tab:coreHrates} for detailed results. Particularly the first five reactions, $^{26}$Al(n,p)$^{26}$Mg, $^{25}$Mg($\alpha$,n)$^{28}$Si, $^{24}$Mg(n,$\gamma$)$^{25}$Mg, $^{23}$Na($\alpha$,p)$^{26}$Mg and $^{26}$Al(n,$\alpha$)$^{23}$Na, should be prime targets for future measurements. The approximate temperature range near which the rate needs to be improved ($\approx 2.3$ GK for explosive Ne/C burning, $\approx 1.4$ GK for convective shell C/Ne burning), as well as the current literature source of a particular rate, is also given in the table. For those five reactions we argued in $\S$ \ref{uncert}, $\S$ \ref{uncertConvCb} and Appendix \ref{specrates}
that the current rate uncertainties at astrophysically important temperatures amount to about a factor of 2. The sensitivity of $^{26}$Al production to rate variations of these reactions can be estimated from Tabs. \ref{tabxNe} and \ref{tabcC}: a factor of 2 variation in their rates changes the final $^{26}$Al mass fraction by factors of 1.7, 1.9, 1.6, 1.3 and 1.3, respectively. Thus we conclude that {\it the uncertainty of the $^{26}$Al yield predicted by the massive star models explored here amounts to about a factor of 3}. This result is obtained on the basis of nuclear physics uncertainties alone and should be considered together with other uncertainties inherent in the stellar models, such as mixing, mass loss and rotation. 

\begin{deluxetable}{lccl}
\tablewidth{0pt}
\tablecaption{Summary of nuclear reactions that impact $^{26}$Al production in massive stars\tablenotemark{a}, assuming thermal equilibrium for $^{26}$Al\label{tabsum}}
\tablehead{
\colhead{Reaction} &  \colhead{Site\tablenotemark{b}}  &  \colhead{Temperature\tablenotemark{c}}   &  \colhead{Source\tablenotemark{d}}   
}
\startdata
$^{26}$Al$^t$(n,p)$^{26}$Mg			     & xNe/C; C/Ne  &  $\approx 2.3$; $\approx 1.4$  &  present    \\
$^{25}$Mg($\alpha$,n)$^{28}$Si		     & xNe/C            &  $\approx 2.3$                          &  nacr    \\
$^{24}$Mg(n,$\gamma$)$^{25}$Mg		     & xNe/C; C/Ne  &  $\approx 2.3$; $\approx 1.4$  &  ka02   \\
$^{23}$Na($\alpha$,p)$^{26}$Mg		     & C/Ne              &  $\approx 1.4$                          &  rath      \\
$^{26}$Al$^t$(n,$\alpha$)$^{23}$Na	     & xNe/C; C/Ne  &  $\approx 2.3$; $\approx 1.4$  &  present     \\
$^{27}$Al($\alpha$,p)$^{30}$Si	             & xNe/C            &  $\approx 2.3$                          &  rath   \\
$^{29}$Si($\alpha$,n)$^{32}$S		             & xNe/C            &  $\approx 2.3$                          &  rath   \\
$^{26}$Mg($\alpha$,n)$^{29}$Si		     & C/Ne              & $\approx 1.4$                           &  nacr     \\
\enddata
\tablenotetext{a}{In approximate order of importance; for full results, see Tabs. \ref{tabxNe}, \ref{tabcC} and \ref{tab:coreHrates}.}
\tablenotetext{b}{Site of $^{26}$Al synthesis in massive star; the labels ``xNe/C" and ``C/Ne" refer to explosive Ne/C burning and convective shell C/Ne burning, respectively.}
\tablenotetext{c}{Temperature (in units of GK) near which most of $^{26}$Al production occurs in given site.}
\tablenotetext{d}{For reference labels, see Tabs. \ref{tabxNe} and \ref{tabcC}.}
\end{deluxetable}

We do not list any reactions for core H burning in Tab. \ref{tab:coreHrates}, mainly because in this case reaction rate variations have only a small effect on the $^{26}$Al yield. Here, the most important reaction is $^{26}$Al(p,$\gamma$)$^{27}$Si, but even a factor of 10 change in this rate near $\approx 90$ MK has only a modest impact on the $^{26}$Al yield (see Tab. \ref{tab:coreHrates}). Of course, new experimental results for $^{26}$Al(p,$\gamma$)$^{27}$Si are useful in any case.

We carefully examined the issue of $^{26}$Al equilibration for each of the three nucleosynthesis sites mentioned above. Two series of post-processing calculations were performed and the resulting $^{26}$Al yields were compared: one assuming either a single or two separate $^{26}$Al species, depending on the temperature regime, and one where the communication between ground and isomeric states was explicitly taken into account. For the latter case, no artificial assumptions about the equilibration of $^{26}$Al are made, but additional $^{26}$Al species (i.e., levels at 417, 1058 and 2070 keV; see Fig. \ref{figlevel}) were taken into account in the reaction network. We found that the equilibration of $^{26}$Al levels in any of the massive star sites investigated here has only minor effects on the $^{26}$Al yields. The reason is that in explosive Ne/C burning and convective shell C/Ne burning the temperatures are sufficiently high to ensure thermal equilibration of $^{26}$Al, while in core H burning the temperatures are never high enough to facilitate communication of the ground and isomeric state via thermal excitations. We also verified that current uncertainties in some unmeasured $^{26}$Al $\gamma$-ray transition rates do not significantly impact the predicted nucleosynthesis yields.

For the interested reader we provide detailed comments on the status of certain reactions, including $^{12}$C($^{12}$C,n)$^{23}$Mg, $^{23}$Na($\alpha$,p)$^{26}$Mg, $^{25}$Mg($\alpha$,n)$^{28}$Si, $^{26}$Al$^{m}$(p,$\gamma$)$^{27}$Si, $^{26}$Al(n,p)$^{26}$Mg and $^{26}$Al(n,$\alpha$)$^{23}$Na. For the latter two, particularly important, reactions we provide new rate estimates, which will be presented in more detail in a forthcoming publication (Oginni et al., in print).

\acknowledgments{This work was supported in part by the U.S. Department of Energy under Contract No. DE-FG02-97ER41041.}

\appendix
\section{$\beta$- AND $\gamma$-DECAY RATES OF $^{26}$AL LEVELS\label{appdec}}
The decay constants for $\beta$- and $\gamma$-decay of $^{26}$Al levels, in units of $s^{-1}$, are listed in Tab. \ref{tblaldecay} for the temperature range of relevance in the present work. The labels $^{26}$Al$^g$, $^{26}$Al$^m$, $^{26}$Al$^a$, $^{26}$Al$^b$ and $^{26}$Al$^c$ refer to the levels at $E_x=0$ keV (5$^+$; ground state), 228 keV (0$^+$; isomeric state), 417 keV (3$^+$), 1058 keV (1$^+$) and 2070 keV (2$^+$), respectively (see Fig. \ref{figlevel}). 

The entries in the second column refer to the $\beta$-decay of $^{26}$Al$^t$ (ground and isomeric state in thermal equilibrium) to the daughter $^{26}$Mg. The decay constant is only listed for temperatures above $T=0.4$ GK since for lower temperatures thermal equilibrium is not achieved. The values are calculated from $\lambda(^{26}$Al$^t$$\rightarrow^{26}$Mg$)=9.93\times10^{-3}~e^{-2.646/T_9}$ s$^{-1}$, where $T_9$ is the temperature in GK (see Ward \& Fowler 1980, Iliadis 2007). This expression, which takes only the ground and isomeric state into account, is valid for temperatures and densities below 5 GK and 10$^6$ g/cm$^{3}$, respectively. In the temperature and density regimes considered here, the results are in good agreement with the more extensive calculations of Oda et al. (1994). Note that the original REACLIB fit of this particular decay rate is off by $\approx$20-80\% at $T=1-3$ GK.

The following $\beta$-decay constants are not listed in the table since they are constant for the temperature grid shown here:\\
$\lambda(^{26}$Al$^g\rightarrow^{26}$Mg)=3.069$\times$10$^{-14}$ $s^{-1}$\\
$\lambda(^{26}$Al$^m\rightarrow^{26}$Mg)=1.092$\times$10$^{-1}$ $s^{-1}$\\
$\lambda(^{26}$Al$^a\rightarrow^{26}$Mg)=1.283$\times$10$^{-4}$ $s^{-1}$\\
$\lambda(^{26}$Al$^b\rightarrow^{26}$Mg)=9.630$\times$10$^{-2}$ $s^{-1}$\\
The first two values are computed from measured laboratory half-lifes (Audi et al. 2003), while the latter two values are obtained from shell model calculations (Kajino et al. 1988).

The following $\gamma$-ray decay constants are not listed in the table since they are nearly constant for the temperature grid shown here:\\
$\lambda(^{26}$Al$^b\rightarrow^{26}$Al$^m$)=2.780$\times$10$^{13}$ $s^{-1}$\\
$\lambda(^{26}$Al$^c\rightarrow^{26}$Al$^m$)=1.500$\times$10$^{12}$ $s^{-1}$\\
$\lambda(^{26}$Al$^b\rightarrow^{26}$Al$^a$)=7.240$\times$10$^{8}$ $s^{-1}$\\
$\lambda(^{26}$Al$^c\rightarrow^{26}$Al$^a$)=1.070$\times$10$^{13}$ $s^{-1}$\\
$\lambda(^{26}$Al$^c\rightarrow^{26}$Al$^b$)=3.770$\times$10$^{13}$ $s^{-1}$\\
All $\gamma$-ray decay constants given above and listed in Tab. \ref{tblaldecay} are calculated from experimental information (Endt 1990), except those connecting the levels $^{26}$Al$^m\leftrightarrow^{26}$Al$^a$ and $^{26}$Al$^a\leftrightarrow^{26}$Al$^b$, which have been found from shell model calculations (Runkle, Champagne \& Engel 2001; see also Fig. \ref{figlevel}).

\clearpage
\begin{deluxetable}{lcccccccccc}
\tabletypesize{\scriptsize}
\rotate
\tablecaption{Decay constants of $^{26}$Al levels (in s$^{-1}$)\label{tblaldecay}}
\tablewidth{0pt}
\tablehead{
\colhead{T (GK)} & \colhead{$^{26}$Al$^t$$\rightarrow^{26}$Mg} & \colhead{$^{26}$Al$^g\rightarrow^{26}$Al$^a$} & \colhead{$^{26}$Al$^a\rightarrow^{26}$Al$^g$} & \colhead{$^{26}$Al$^m\rightarrow^{26}$Al$^a$} &
\colhead{$^{26}$Al$^a\rightarrow^{26}$Al$^m$} & \colhead{$^{26}$Al$^m\rightarrow^{26}$Al$^b$} & \colhead{$^{26}$Al$^m\rightarrow^{26}$Al$^c$} &
\colhead{$^{26}$Al$^a\rightarrow^{26}$Al$^b$} & \colhead{$^{26}$Al$^a\rightarrow^{26}$Al$^c$} & \colhead{$^{26}$Al$^b\rightarrow^{26}$Al$^c$}
}
\startdata
0.04	& \nodata  	& 1.12E-44	& 5.56E+08	& 7.11E-25	& 6.24E-02	&  \nodata 	&  \nodata 	&  \nodata 	&  \nodata 	&  \nodata  \\
0.05	& \nodata  	& 3.43E-34	& 5.56E+08	& 4.07E-20	& 6.24E-02	&  \nodata 	&  \nodata 	&  \nodata 	&  \nodata 	&  \nodata  \\
0.06	& \nodata  	& 3.45E-27	& 5.56E+08	& 6.05E-17	& 6.24E-02	&  \nodata 	&  \nodata 	&  \nodata 	&  \nodata 	&  \nodata  \\
0.07	& \nodata  	& 3.46E-22	& 5.56E+08	& 1.11E-14	& 6.24E-02	&  \nodata 	&  \nodata 	& 2.21E-38	&  \nodata 	&  \nodata  \\
0.08	& \nodata  	& 1.95E-18	& 5.56E+08	& 5.58E-13	& 6.24E-02	& 4.45E-39	&  \nodata 	& 1.30E-32	&  \nodata 	&  \nodata  \\
0.09	& \nodata  	& 1.62E-15	& 5.56E+08	& 1.17E-11	& 6.24E-02	& 2.86E-33	&  \nodata 	& 3.97E-28	&  \nodata 	& 1.37E-43 \\
0.10	& \nodata  	& 3.49E-13	& 5.56E+08	& 1.33E-10	& 6.24E-02	& 1.27E-28	&  \nodata 	& 1.54E-24	&  \nodata 	& 6.36E-38 \\
0.11	& \nodata  	& 2.83E-11	& 5.56E+08	& 9.77E-10	& 6.24E-02	& 8.03E-25	&  \nodata 	& 1.33E-21	&  \nodata 	& 2.75E-33 \\
0.12	& \nodata  	& 1.10E-09	& 5.56E+08	& 5.14E-09	& 6.24E-02	& 1.18E-21	&  \nodata 	& 3.73E-19	&  \nodata 	& 2.01E-29 \\
0.13	& \nodata  	& 2.45E-08	& 5.56E+08	& 2.09E-08	& 6.24E-02	& 5.67E-19	&  \nodata 	& 4.40E-17	&  \nodata 	& 3.72E-26 \\
0.14	& \nodata  	& 3.50E-07	& 5.56E+08	& 6.98E-08	& 6.24E-02	& 1.13E-16	&  \nodata 	& 2.62E-15	&  \nodata 	& 2.36E-23 \\
0.15	& \nodata  	& 3.50E-06	& 5.56E+08	& 1.98E-07	& 6.24E-02	& 1.10E-14	&  \nodata 	& 9.04E-14	& 2.24E-43	& 6.33E-21 \\
0.16	& \nodata  	& 2.63E-05	& 5.56E+08	& 4.94E-07	& 6.24E-02	& 6.09E-13	& 1.40E-45	& 2.00E-12	& 6.64E-40	& 8.44E-19 \\
0.18	& \nodata  	& 7.56E-04	& 5.56E+08	& 2.26E-06	& 6.24E-02	& 4.89E-10	& 2.05E-39	& 3.51E-10	& 4.04E-34	& 2.93E-15 \\
0.20	& \nodata  	& 1.10E-02	& 5.56E+08	& 7.63E-06	& 6.24E-02	& 1.03E-07	& 2.94E-34	& 2.19E-08	& 1.71E-29	& 2.00E-12 \\
0.25	& \nodata  	& 1.40E+00	& 5.56E+08	& 6.83E-05	& 6.24E-02	& 1.56E-03	& 5.62E-25	& 3.72E-05	& 3.66E-21	& 2.51E-07 \\
0.30	& \nodata  	& 3.52E+01	& 5.56E+08	& 2.94E-04	& 6.24E-02	& 9.59E-01	& 8.66E-19	& 5.29E-03	& 1.31E-15	& 6.31E-04 \\
0.35	& \nodata  	& 3.52E+02	& 5.56E+08	& 8.37E-04	& 6.25E-02	& 9.40E+01	& 2.28E-14	& 1.83E-01	& 1.21E-11	& 1.69E-01 \\
0.40	& \nodata  	& 1.98E+03	& 5.56E+08	& 1.83E-03	& 6.27E-02	& 2.93E+03	& 4.70E-11	& 2.60E+00	& 1.15E-08	& 1.12E+01 \\
0.45	& 2.78E-05	& 7.60E+03	& 5.56E+08	& 3.38E-03	& 6.29E-02	& 4.25E+04	& 1.78E-08	& 2.06E+01	& 2.36E-06	& 2.92E+02 \\
0.50	& 5.00E-05	& 2.23E+04	& 5.56E+08	& 5.53E-03	& 6.32E-02	& 3.61E+05	& 2.05E-06	& 1.07E+02	& 1.67E-04	& 3.97E+03 \\
0.60	& 1.21E-04	& 1.12E+05	& 5.56E+08	& 1.20E-02	& 6.41E-02	& 8.94E+06	& 2.55E-03	& 1.28E+03	& 1.00E-01	& 1.99E+05 \\
0.70	& 2.27E-04	& 3.54E+05	& 5.57E+08	& 2.00E-02	& 6.53E-02	& 8.85E+07	& 4.13E-01	& 7.53E+03	& 9.63E+00	& 3.26E+06 \\
0.80	& 3.64E-04	& 8.40E+05	& 5.57E+08	& 3.00E-02	& 6.67E-02	& 4.94E+08	& 1.88E+01	& 2.84E+04	& 2.96E+02	& 2.65E+07 \\
0.90	& 5.25E-04	& 1.65E+06	& 5.59E+08	& 4.20E-02	& 6.84E-02	& 1.88E+09	& 3.65E+02	& 7.98E+04	& 4.24E+03	& 1.35E+08 \\
1.00	& 7.04E-04	& 2.83E+06	& 5.60E+08	& 5.50E-02	& 7.02E-02	& 5.49E+09	& 3.92E+03	& 1.82E+05	& 3.58E+04	& 4.99E+08 \\
1.25	& 1.20E-03	& 7.54E+06	& 5.68E+08	& 9.20E-02	& 7.55E-02	& 3.77E+10	& 2.82E+05	& 8.10E+05	& 1.66E+06	& 5.23E+09 \\
1.50	& 1.70E-03	& 1.47E+07	& 5.79E+08	& 1.32E-01	& 8.13E-02	& 1.36E+11	& 4.87E+06	& 2.19E+06	& 2.14E+07	& 2.50E+10 \\
1.75	& 2.19E-03	& 2.38E+07	& 5.93E+08	& 1.75E-01	& 8.74E-02	& 3.42E+11	& 3.73E+07	& 4.49E+06	& 1.33E+08	& 7.66E+10 \\
2.00	& 2.64E-03	& 3.46E+07	& 6.10E+08	& 2.20E-01	& 9.38E-02	& 6.82E+11	& 1.72E+08	& 7.71E+06	& 5.23E+08	& 1.78E+11 \\
2.50	& 3.45E-03	& 5.98E+07	& 6.50E+08	& 3.12E-01	& 1.07E-01	& 1.81E+12	& 1.45E+09	& 1.67E+07	& 3.56E+09	& 5.78E+11 \\
\enddata
\tablecomments{Table \ref{tblaldecay} is published in the electronic edition of the {\it Astrophysical Journal}.}
\end{deluxetable}

\section{REACTION RATES INVOLVING $^{26}$AL$^t$, $^{26}$AL$^g$ and $^{26}$AL$^m$\label{appreac}}
Information on the reaction rates involving the production and destruction of $^{26}$Al species in our network is given in Tab. \ref{tblalrates}. The second row lists the source of the rates: ``il10" (experimental Monte Carlo rates from Iliadis et al. 2010); ``rath" (theoretical Hauser-Feshbach rates from Rauscher \& Thielemann 2000); ``present" (hybrid rate, see below). We argued in $\S$ \ref{thermequi} that it is important to ensure internal consistency of the rates used. For example, for the first three listed reactions ($^{25}$Mg+p) the rates are based on the same nuclear physics input and are thus consistent. Similar arguments apply to the following two reactions ($^{26}$Al$^x$+p). 

However, the situation for the $^{26}$Al$^m$(p,$\gamma$)$^{27}$Si reaction is a different matter. Rates have been estimated in Caughlan \& Fowler (1988) and in Angulo et al. (1999), while initial experimental studies are reported in Deibel et al. (2009) and Lotay et al. (2009). There can be no doubt that this rate is highly uncertain at present (see Appendix \ref{al26mpgapp}). In the absence of a better procedure, we approximated the $^{26}$Al$^m$(p,$\gamma$)$^{27}$Si rate by the (experimental) ground state rate (column 6 in Tab. \ref{tblalrates}). Note that these two rates were predicted by Caughlan \& Fowler (1988) to be similar within a factor of $\approx$5. Our assumption should be regarded as a starting point for exploring the effects of $^{26}$Al$^m$(p,$\gamma$)$^{27}$Si reaction rate variations.

For the $^{26}$Al$^t$(n,$\alpha$)$^{23}$Na reaction we use a hybrid rate, which is based on experimental information from De Smet et al. (2007) at $T\leq0.1$ GK, and on Hauser-Feshbach results from Rauscher \& Thielemann (2000) at higher temperatures (see Appendix \ref{al26na}; the rate is listed in the last column of Tab. \ref{tblalrates}). The predicted stellar enhancement factors are relatively small (43\% at 2.5 GK) and, therefore, we adopt these rates also for the $^{26}$Al$^g$(n,$\alpha$)$^{23}$Na reaction. Furthermore, in order to ensure internal consistency, we approximated the $^{26}$Al$^m$(n,$\alpha$)$^{23}$Na rate by the thermalized rate. Note that the thermalized and isomeric state rates were predicted by Caughlan \& Fowler (1988) to be similar within a factor of $\approx$7. Again, our assumptions serve as starting points to explore the effects of $^{26}$Al$^x$(n,$\alpha$)$^{23}$Na reaction rate variations. A similar procedure has been followed for the $^{26}$Al$^x$(n,p)$^{26}$Mg reaction rates (see column 7 of Tab. \ref{tblalrates} and Appendix \ref{al26np}).

The rates of the following reactions are not explicitly listed in Tab. \ref{tblalrates}: $^{22}$Na($\alpha$,$\gamma$)$^{26}$Al$^t$, $^{25}$Al(n,$\gamma$)$^{26}$Al$^t$, $^{23}$Mg($\alpha$,p)$^{26}$Al$^t$, $^{26}$Si(n,p)$^{26}$Al$^t$, $^{29}$P(n,$\alpha$)$^{26}$Al$^t$, $^{26}$Al$^t$(n,$\gamma$)$^{27}$Al, $^{26}$Al$^t$($\alpha$,$\gamma$)$^{30}$P and $^{26}$Al$^t$($\alpha$,p)$^{29}$Si. For these we used the Hauser-Feshbach rates of Rauscher \& Thielemann (2000), except the rates of $^{26}$Al$^t$(n,$\gamma$)$^{27}$Al, which were adopted from the  KADoNiS v0.2 evaluation (Dillmann et al. 2006). Since the stellar enhancement factors are predicted to be small, we also adopted these results for the respective rates involving $^{26}$Al$^g$. Note that the corresponding reactions involving $^{26}$Al$^m$ are absent in the original REACLIB. We disregarded these as well, on the grounds that their net abundance flows (for $^{26}$Al$^t$) in our network calculations are at least 3 orders of magnitude smaller than the maximum flow.

For all forward reactions discussed above, the corresponding reverse reaction rates are also implemented in our library. Note that in these cases we estimated the reverse rates using the principle of detailed balance assuming a single (ground or isomeric) level in $^{26}$Al only. The proper procedure would have been to apply detailed balance to the forward rate involving the quasi-equilibrium cluster of levels in thermal equilibrium with the $^{26}$Al level in question. We are not aware that this information has been given anywhere before and believe that the effects of our approximation are very small.

\clearpage
\begin{deluxetable}{lccccccc}
\tabletypesize{\scriptsize}
\rotate
\tablecaption{Stellar reaction rates\tablenotemark{a} $N_A\left<\sigma v \right>$ involving $^{26}$Al (in cm$^3$~mol$^{-1}$~s$^{-1}$)\label{tblalrates}}
\tablewidth{0pt}
\tablehead{
\colhead{T (GK)} & \colhead{$^{25}$Mg(p,$\gamma$)$^{26}$Al$^t$} & \colhead{$^{25}$Mg(p,$\gamma$)$^{26}$Al$^g$} & \colhead{$^{25}$Mg(p,$\gamma$)$^{26}$Al$^m$} & \colhead{$^{26}$Al$^t$(p,$\gamma$)$^{27}$Si} &
\colhead{$^{26}$Al$^g$(p,$\gamma$)$^{27}$Si} & \colhead{$^{26}$Al$^t$(n,p)$^{26}$Mg} & \colhead{$^{26}$Al$^t$(n,$\alpha$)$^{23}$Na} \\
 & \colhead{il10\tablenotemark{b}} & \colhead{il10\tablenotemark{b}} & \colhead{il10\tablenotemark{b}} & \colhead{il10\tablenotemark{b}} & \colhead{il10\tablenotemark{b,c}} & \colhead{present\tablenotemark{d,f}} & \colhead{present\tablenotemark{e,f}} 
}
\startdata
0.015 &	1.07e-24 &	8.71e-25 &	2.04e-25 &	4.07e-31 &	4.07e-31 &	2.98e+06 &	4.39e+06 \\	
0.016 &	1.56e-23 &	1.27e-23 &	2.96e-24 &	4.57e-30 &	4.57e-30 &	3.40e+06 &	5.20e+06 \\	
0.018 &	1.35e-21 &	1.09e-21 &	2.55e-22 &	6.05e-28 &	6.05e-28 &	4.19e+06 &	6.77e+06 \\	
0.020 &	4.69e-20 &	3.81e-20 &	8.89e-21 &	4.78e-26 &	4.78e-26 &	4.90e+06 &	8.17e+06 \\	
0.025 &	2.67e-17 &	2.16e-17 &	5.05e-18 &	2.73e-22 &	2.73e-22 &	6.37e+06 &	1.09e+07 \\	
0.030 &	1.74e-15 &	1.41e-15 &	3.30e-16 &	1.49e-19 &	1.49e-19 &	7.53e+06 &	1.25e+07 \\	
0.040 &	3.00e-13 &	2.44e-13 &	5.68e-14 &	7.35e-16 &	7.35e-16 &	9.41e+06 &	1.39e+07 \\	
0.050 &	6.82e-12 &	5.57e-12 &	1.26e-12 &	1.25e-13 &	1.25e-13 &	1.11e+07 &	1.41e+07 \\	
0.060 &	6.56e-11 &	5.43e-11 &	1.15e-11 &	3.91e-12 &	3.91e-12 &	1.27e+07 &	1.39e+07 \\	
0.070 &	3.99e-10 &	3.33e-10 &	6.67e-11 &	5.71e-11 &	5.71e-11 &	1.42e+07 &	1.37e+07 \\	
0.080 &	1.72e-09 &	1.45e-09 &	2.81e-10 &	7.80e-10 &	7.80e-10 &	1.58e+07 &	1.34e+07 \\	
0.090 &	5.73e-09 &	4.85e-09 &	9.35e-10 &	9.54e-09 &	9.54e-09 &	1.74e+07 &	1.31e+07 \\	
0.100 &	1.60e-08 &	1.34e-08 &	2.72e-09 &	8.17e-08 &	8.17e-08 &	1.91e+07 &	1.28e+07 \\	
0.110 &	4.14e-08 &	3.40e-08 &	7.78e-09 &	4.94e-07 &	4.94e-07 &	2.07e+07 &	1.26e+07 \\	
0.120 &	1.17e-07 &	9.36e-08 &	2.38e-08 &	2.24e-06 &	2.24e-06 &	2.24e+07 &	1.24e+07 \\	
0.130 &	4.05e-07 &	3.25e-07 &	8.12e-08 &	8.10e-06 &	8.10e-06 &	2.41e+07 &	1.23e+07 \\	
0.140 &	1.62e-06 &	1.32e-06 &	2.93e-07 &	2.45e-05 &	2.45e-05 &	2.58e+07 &	1.21e+07 \\	
0.150 &	6.40e-06 &	5.34e-06 &	1.05e-06 &	6.41e-05 &	6.41e-05 &	2.74e+07 &	1.20e+07 \\	
0.160 &	2.31e-05 &	1.95e-05 &	3.53e-06 &	1.50e-04 &	1.50e-04 &	2.91e+07 &	1.19e+07 \\	
0.180 &	2.10e-04 &	1.79e-04 &	3.02e-05 &	6.38e-04 &	6.38e-04 &	3.24e+07 &	1.18e+07	 \\
0.200 &	1.26e-03 &	1.08e-03 &	1.79e-04 &	2.13e-03 &	2.13e-03 &	3.57e+07 &	1.17e+07	 \\
0.250 &	3.20e-02 &	2.73e-02 &	4.68e-03 &	2.30e-02 &	2.30e-02 &	4.36e+07 &	1.17e+07	 \\
0.300 &	2.76e-01 &	2.33e-01 &	4.25e-02 &	1.40e-01 &	1.40e-01 &	5.10e+07 &	1.17e+07	 \\
0.350 &	1.29e+00 &	1.08e+00 &	2.10e-01 &	5.74e-01 &	5.74e-01 &	5.77e+07 &	1.19e+07	 \\
0.400 &	4.16e+00 &	3.45e+00 &	7.06e-01 &	1.74e+00 &	1.74e+00 &	6.38e+07 &	1.21e+07	 \\
0.450 &	1.04e+01 &	8.55e+00 &	1.85e+00 &	4.19e+00 &	4.19e+00 &	6.93e+07 &	1.23e+07	 \\
0.500 &	2.19e+01 &	1.78e+01 &	4.06e+00 &	8.50e+00 &	8.50e+00 &	7.43e+07 &	1.26e+07	 \\
0.600 &	6.84e+01 &	5.46e+01 &	1.37e+01 &	2.45e+01 &	2.45e+01 &	8.27e+07 &	1.32e+07	 \\
0.700 &	1.58e+02 &	1.24e+02 &	3.38e+01 &	5.14e+01 &	5.14e+01 &	8.95e+07 &	1.39e+07	 \\
0.800 &	2.99e+02 &	2.31e+02 &	6.80e+01 &	8.96e+01 &	8.87e+01 &	9.51e+07 &	1.46e+07	 \\
0.900 &	4.98e+02 &	3.78e+02 &	1.19e+02 &	1.36e+02 &	1.35e+02 &	9.98e+07 &	1.53e+07	 \\
1.000 &	7.53e+02 &	5.65e+02 &	1.88e+02 &	1.92e+02 &	1.88e+02 &	1.04e+08 &	1.61e+07	 \\
1.250 &	1.62e+03 &	1.18e+03 &	4.38e+02 &	3.59e+02 &	3.45e+02 &	1.12e+08 &	1.82e+07	 \\
1.500 &	2.75e+03 &	1.98e+03 &	7.88e+02 &	5.57e+02 &	5.25e+02 &	1.19e+08 &	2.04e+07	 \\
1.750 &	4.09e+03 &	2.89e+03 &	1.22e+03 &	7.84e+02 &	7.19e+02 &	1.25e+08 &	2.29e+07	 \\
2.000 &	5.56e+03 &	3.90e+03 &	1.71e+03 &	1.14e+03 &	1.01e+03 &	1.31e+08 &	2.55e+07	 \\
2.500 &	8.66e+03 &	6.08e+03 &	2.79e+03 &	1.99e+03 &	1.67e+03 &	1.46e+08 &	3.14e+07	 \\
\enddata
\tablenotetext{a}{The rates for only some reactions are listed; other reactions are discussed in the text. All rates given here (except column 6) account for thermal target excitations.}
\tablenotetext{b}{Experimental Monte Carlo rates of Iliadis et al. (2010).}
\tablenotetext{c}{Same rate is used for $^{26}$Al$^m$(p,$\gamma$)$^{27}$Si.}
\tablenotetext{d}{Hybrid rate: at $T\leq0.2$ GK from experiment of Koehler et al. (1997); at $T>0.2$ GK from Hauser-Feshbach model of Rauscher \& Thielemann (2000).}
\tablenotetext{e}{Hybrid rate: at $T\leq0.1$ GK from experiment of De Smet et al. (2007); at $T>0.1$ GK from Hauser-Feshbach model of Rauscher \& Thielemann (2000).}
\tablenotetext{f}{Same rate is used for $^{26}$Al$^g$ and $^{26}$Al$^m$.}
\tablecomments{Table \ref{tblalrates} is published in the electronic edition of the {\it Astrophysical Journal}.}
\end{deluxetable}

\section{DISCUSSION OF SPECIFIC REACTION RATES\label{specrates}}
The references of the reaction rates used in the present work are provided in the tables above and the reader is referred to these sources for details. Decays and reactions involving levels of $^{26}$Al are discussed in Appendices \ref{appdec} and \ref{appreac}, respectively. For some specific reaction rates that are neither evaluated in Iliadis et al. (2010) nor in Angulo et al. (1999), we provide more information below.

\subsection{$^{12}$C($^{12}$C,n)$^{23}$Mg ($Q=-2.598$ MeV)\label{c12c12n}}
We calculated the reaction rate from the total $^{12}$C+$^{12}$C rate (i.e., summed over all exit channels) and the neutron branching ratio. For $T=1.25$ GK, representing the average temperature of convective shell C/Ne burning of the 60 $M_\odot$ model explored in the present work, the Gamow peak extends over a center-of-mass energy range from 2.6 MeV (i.e., the threshold energy for this endothermic reaction) to about 3.4 MeV. In this energy region the total $^{12}$C+$^{12}$C cross section has been measured (Costantini et al. 2009, and references therein). For the neutron branching ratio, as a function of temperature in the range of $T=0.5-5.0$ GK, we adopted the values from Dayras, Switkowski and Woosley (1977). In the latter work, the $^{12}$C($^{12}$C,n)$^{23}$Mg reaction has been measured down to an energy of 3.54 MeV and, with the aid of statistical model calculations, the results were extrapolated down to the threshold energy in order to extract the neutron branching ratio. The overall reaction rate uncertainty is difficult to quantify at present. Clearly, a new measurement of the $^{12}$C($^{12}$C,n)$^{23}$Mg reaction at lower energies is desirable.
 
The $^{12}$C($^{12}$C,n)$^{23}$Mg reaction rates employed in various rate libraries are inconsistent with each other. An impression can be obtained from Fig. \ref{figc12c12}. The black line represents the ratio of the rate used in the Basel version of the REACLIB (nucastro.org/reaclib.html\#reaclib) to the present rate. The hatched area marks the temperature range of convective shell C/Ne burning explored in the present work. Surprisingly, the ratio amounts to more than an order of magnitude near $T=1.25$ GK. The reason may be that the $^{12}$C($^{12}$C,n)$^{23}$Mg rate in the original REACLIB library was erroneously obtained using a neutron branching ratio of zero for $T<1.75$ GK that is listed in Caughlan \& Fowler (1988). The blue line in Fig. \ref{figc12c12} represents the ratio of the rate used in the MSU version of the REACLIB (groups.nscl.msu.edu/jina/reaclib/db/) to the present rate. The ratio amounts to a factor of 2 near $T=1.25$ GK. The reason may perhaps be that their fitted rate deviates from the actual rate. For the interested reader, we provide below a rate fit (in REACLIB format) that reproduces the present and best currently available rate within 10\%:
\begin{verbatim}
-0.527166E+02
-0.344948E+02 
 0.140849E+03
-0.878184E+02
 0.371377E+01
-0.338673E+00
 0.736030E+02
\end{verbatim}

\begin{figure}
\includegraphics[scale=0.5]{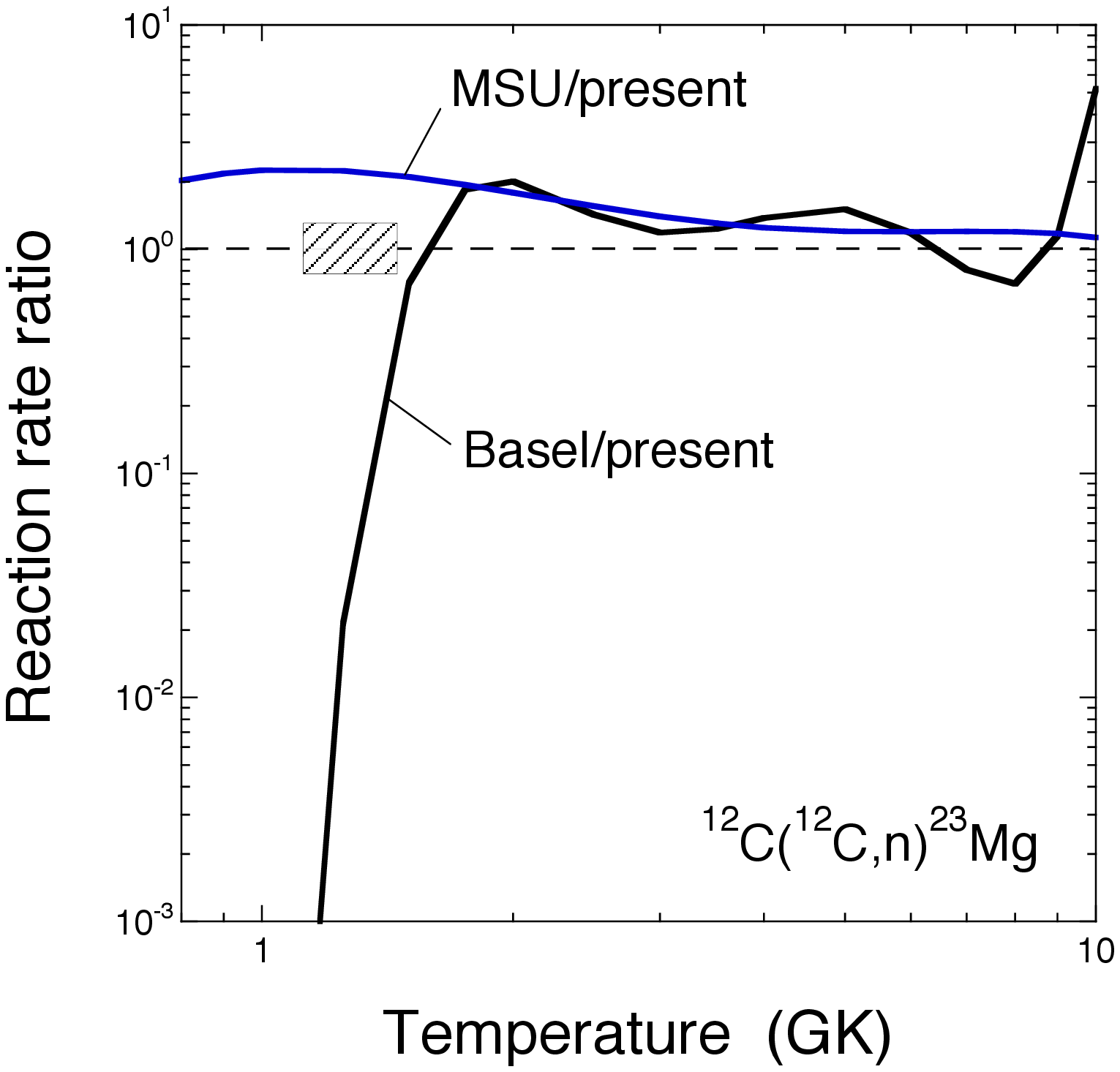}
\caption{Reaction rate ratio for $^{12}$C($^{12}$C,n)$^{23}$Mg: (black line) ratio of ``Basel" rate to present rate; (blue line) ratio of ``MSU" rate to present rate. The hatched region marks the temperature range of convective shell C/Ne burning explored in the present work. \label{figc12c12}}
\end{figure}

\subsection{$^{23}$Na($\alpha$,p)$^{26}$Mg ($Q=1.821$ MeV)\label{na23apapp}}
A direct measurement of this reaction has been reported in Whitmire \& Davids (1974). They bombarded a target, fabricated by evaporating NaCl onto a thick Cu backing, with $\alpha$-particles (E$_{cm}=2.0-3.1$ MeV) and measured the emitted protons populating the ground and first excited states in $^{26}$Mg. In total, they report the strengths of 39 resonances. There are a number of reasons that warrant a re-measurement of this reaction. Most importantly, the strengths have been determined relative to an absolute strength measurement for the E$_{lab}=3051$ keV resonance, assuming that the stoichiometry of their NaCl target amounts to 1:1. This issue has been discussed in detail by Rowland et al. (2002) in connection with the $^{23}$Na(p,$\alpha$)$^{20}$Ne reaction, where it was shown that during proton bombardment a NaCl target quickly changes its stoichiometry to 5:3, resulting in a significant change in thick-target yield (and in the derived resonance strength). This problem is certainly aggravated when using an $\alpha$-particle beam incident on a NaCl target. There were other problems in the analysis of Whitmire \& Davids (1974): (i) protons populating higher-lying final states could only be resolved at the highest measured energies; (ii) the uncertainty in the resonance energies is relatively large, amounting to $\approx$10 keV; (iii) the assumed uncertainties of 1\% for the stopping cross sections of $\alpha$-particles in Na and Cl seem unreasonably small. Furthermore, for temperatures typical of convective shell C/Ne burning ($T\approx1.25$ GK) the Gamow peak is covering a center-of-mass energy range of $E_{cm}=1.2-2.2$ MeV, i.e., significantly below the energy range covered by experiment. Considering the substantial uncertainties involved, we prefer to use for the reaction rate the estimate provided by the Hauser-Feshbach model (Rauscher \& Thielemann 2000). Clearly, an improved measurement of this reaction is called for. 

\subsection{$^{25}$Mg($\alpha$,n)$^{28}$Si ($Q=2.654$ MeV)\label{mg25anapp}}
Direct measurements of the $^{25}$Mg($\alpha$,n)$^{28}$Si and $^{25}$Mg($\alpha$,n$\gamma$)$^{28}$Si reactions have been reported in Van der Zwan and Geiger (1981), Anderson et al. (1983) and Wieland (1995). The reaction rate recommended by the NACRE collaboration (Angulo et al. 1999) is exclusively based on these data sets. The NACRE collaboration also reports the experimental S-factors for the three references. However, below an energy of 3 MeV the data from Van der Zwan and Geiger (1981) and from Anderson et al. (1983) have been disregarded by NACRE. As a result, in this energy range, their rate is based exclusively on the unpublished work of Wieland (1995), since it is argued in Angulo et al. (1999) that at lower energy background contributions in the earlier works dominate the neutron yield. This conclusion is only partially correct since, for example, Anderson et al. (1983) have also measured the $^{25}$Mg($\alpha$,n$\gamma$)$^{28}$Si reaction, which shows much less background compared to the ($\alpha$,n) reaction. 

\begin{figure}
\includegraphics[scale=0.5]{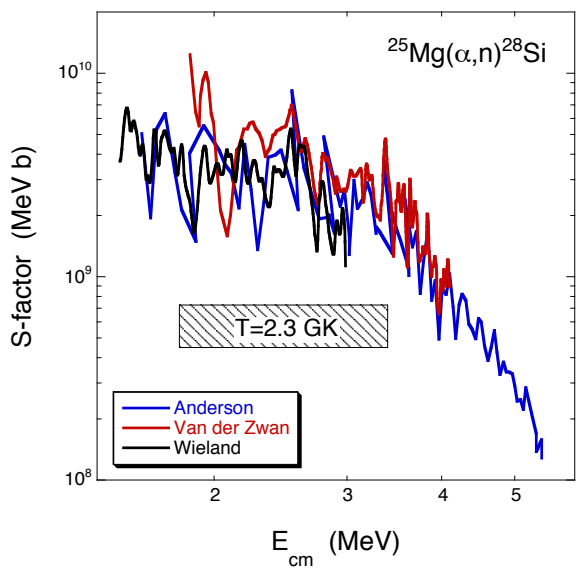}
\caption{Experimental astrophysical S-factors for $^{25}$Mg($\alpha$,n)$^{28}$Si: (red) Van der Zwan and Geiger (1981); (blue) Anderson et al. (1983); (black) Wieland (1995). For the first two references, the data were extracted from the published figures and converted from cross sections to S-factors. For the latter reference, the S-factors are adopted from the NACRE evaluation (Angulo et al. 1999). Uncertainty bars have been omitted for reasons of clarity. The hatched horizontal bar marks the region of the Gamow peak near $T=2.3$ GK, the peak temperature achieved in explosive Ne/C burning.\label{figmg25an}}
\end{figure}

The current situation is shown in Fig. \ref{figmg25an}. The data from Van der Zwan and Geiger (1981) and from Anderson et al. (1983) have been extracted as cross sections from the original figures and converted to astrophysical S-factors. The data from Wieland (1995) are adopted from Angulo et al. (1999). It is interesting to note that the data of Anderson et al. (1983) {\it as reported by the NACRE collaboration} disagree with the corresponding curve shown in Fig. \ref{figmg25an} by a factor of $\approx$2. The reason is presumably that NACRE extracted the ($\alpha$,n) data (which represent only upper limits, as explicitly stated in Anderson et al. 1983) instead of the ($\alpha$,n$\gamma$) data (which are much less susceptible to background). We conclude from Fig. \ref{figmg25an} that the available data are in reasonable agreement. 

It is also inexplicable why already ``...above $T_9=2$, H[auser]F[eshbach] rates are used..." in the NACRE evaluation. The Gamow peak region at $T=2.3$ GK is shown in Fig. \ref{figmg25an} as a hatched horizontal bar. Clearly, the Gamow peak region at this temperature, in fact, up to $T=4$ GK, is entirely covered by experimental data. Thus, there is no reason to use Hauser-Feshbach rates and thereby introduce another source of uncertainty. We feel that a proper re-analysis of the existing cross section data will not only improve the rate of the $^{25}$Mg($\alpha$,n)$^{28}$Si reaction, but will likely provide better estimates of the rate uncertainty, at least up to temperatures of 4 GK. Such an analysis is left for future work. We would also like to reiterate that a new measurement of this reaction would be important.

\subsection{$^{26}$Al$^m$(p,$\gamma$)$^{27}$Si ($Q=7.691$ MeV)\label{al26mpgapp}}
Rates for this reaction are listed in Caughlan \& Fowler (1988), but it is not apparent from their work how the results have been obtained. Presumably their rates were estimated using statistical model calculations (see comments in Ward \& Fowler 1980). In Angulo et al. (1999), the rate for this reaction was obtained by multiplying the experimental (ground state) rate for $^{26}$Al$^g$(p,$\gamma$)$^{27}$Si by the ratio of isomeric and ground state rates, $N_A\left<\sigma v \right>_m$/$N_A\left<\sigma v \right>_g$. The latter ratio was obtained from the Hauser-Feshbach model. However, it is clear from our comments in $\S$ \ref{uncert} that the latter reaction model may not be applicable to $^{26}$Al+p. Furthermore, the rate in Angulo et al. (1999) is only listed at temperatures of $T=0.018-0.4$ GK, but the rate needs to be known at higher temperatures as well in order to study the equilibration of $^{26}$Al levels ($\S$ \ref{thermequi}). 

Recently, some new experimental information has been reported by Deibel et al. (2009) and Lotay et al. (2009). In the former work, levels in the $^{27}$Si compound nucleus near the proton threshold were populated in ($^{3}$He,t) and ($^{3}$He,$\alpha$) reaction studies and the subsequent proton decay to the isomeric state was observed in coincidence, providing values for excitation energies and proton branching ratios. In the latter work, the $^{12}$C($^{16}$O,n) reaction was used to measure $\gamma$-ray transitions in $^{27}$Si, allowing for a determination of excitation energies, $J^\pi$-values and level lifetimes. Nevertheless, too much experimental information is still lacking (i.e., missing levels, spectroscopic factors, proton partial widths, and resonance strengths) in order to estimate this rate reliably over the temperature range of interest. More measurements are clearly in order. 

In the absence of a more reliable estimate, we approximated in this work the $^{26}$Al$^m$(p,$\gamma$)$^{27}$Si rate by the (experimental) ground state rate (see comments in Appendix \ref{appreac}). Our assumption is a starting point for exploring the effects of $^{26}$Al$^m$(p,$\gamma$)$^{27}$Si reaction rate variations.

\subsection{$^{26}$Al(n,$\alpha$)$^{23}$Na ($Q=2.966$ MeV)\label{al26na}}
\begin{figure}
\includegraphics[scale=0.5]{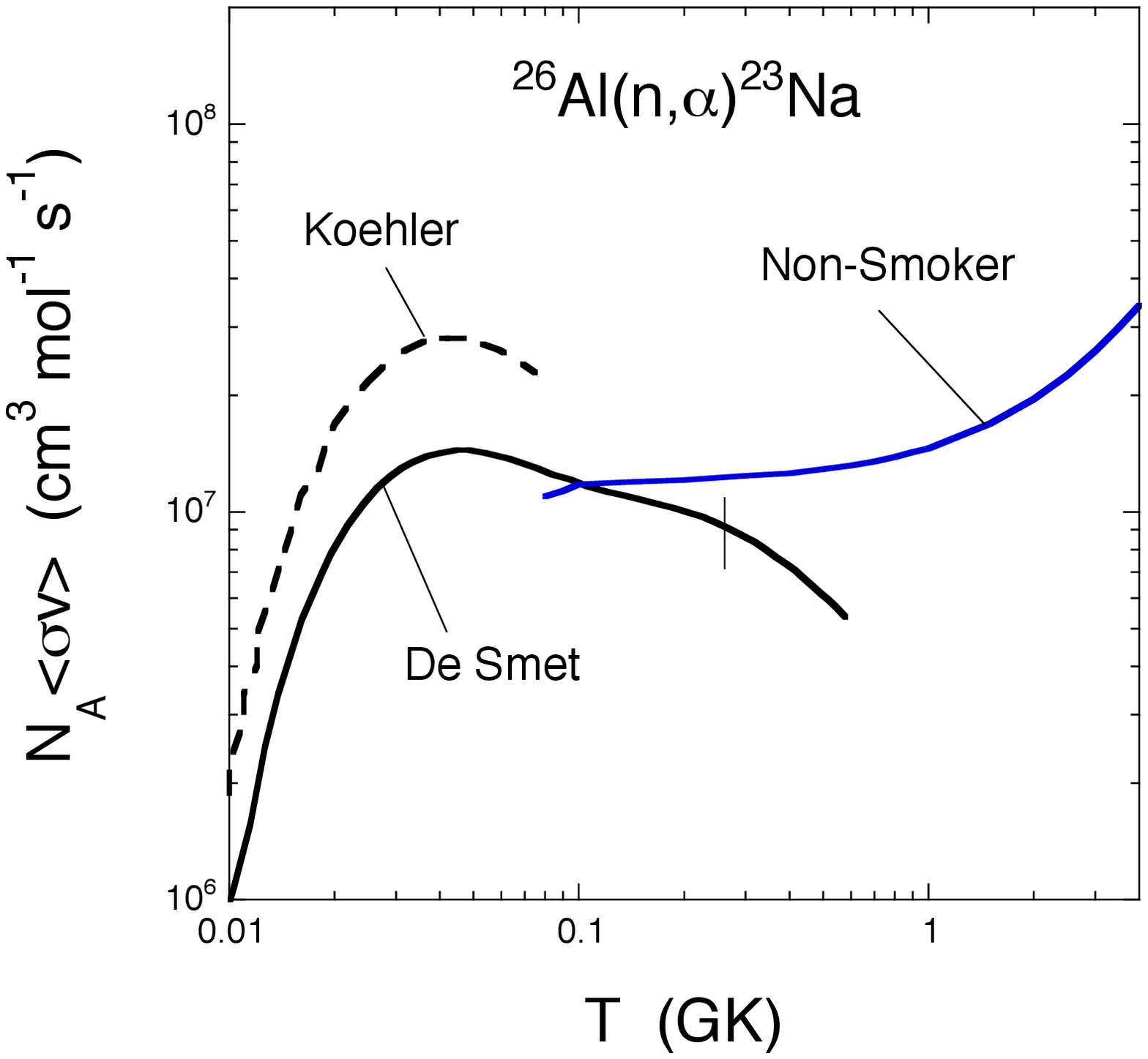}
\caption{Reaction rates for $^{26}$Al(n,$\alpha$)$^{23}$Na: (black solid line) De Smet et al. (2007); (dashed line) Koehler et al. (1997); (blue solid line) Rauscher \& Thielemann (2000). The first two rates are based on experimental results, while the latter rate is estimated using the Hauser-Feshbach model. Beyond the vertical line, near $T\approx0.26$ GK, the experimental rate of De Smet et al. (2007) represents a lower limit. Note that for this comparison only, the rates represent ``laboratory rates", i.e., they do not account for thermal target excitations. \label{figal26na}}
\end{figure}

A direct measurement of the $^{26}$Al(n,$\alpha_0$)$^{23}$Na reaction (i.e., for population of the $^{23}$Na ground sate) has been reported by Koehler et al. (1997), while De Smet et al. (2007) have measured the $^{26}$Al(n,$\alpha_0+\alpha_1$)$^{23}$Na reaction (i.e., for population of the ground and first excited state in $^{23}$Na). The current situation for the reaction rates is displayed in Fig. \ref{figal26na}. The experimental rate for $^{26}$Al(n,$\alpha_0$)$^{23}$Na from Koehler et al. (1997) is extracted from their Fig. 4 and is shown as a dashed line. Note that their rates are claimed to be reliable only for $T\leq0.08$ GK. The more recent experimental rate of De Smet et al. (2007) is shown as a black solid line and was obtained by converting the Maxwellian-averaged cross sections (MACS), shown in their Fig. 8, to reaction rates. This rate represents a {\it lower limit} above $T=0.26$ GK (indicated by the vertical line). The theoretical rate, based on the Hauser-Feshbach model, is adopted from Rauscher \& Thielemann 2000 and is displayed as a solid blue line. A similar theoretical rate has been reported by Goriely, Hilaire \& Koning (2008).

We may draw a number of conclusions from the figure. First, the two experimental rates do not agree, even if the large rate uncertainty (26\%) in the earlier work is taken into account (see discussion in De Smet et al. 2007 for the possible source of the discrepancy). Second, taking both the experimental uncertainty of the De Smet rate into account, as well as the fact that their rate represents a lower limit near their high-temperature cutoff, the agreement with the Hauser-Feshbach rate near $T\approx0.3$ GK is reasonable. Recall that for the purposes of the present work the rate is of interest at temperatures between 1.1 GK (convective shell C/Ne burning) and 2.3 GK (explosive Ne/C burning), i.e., in a region that has not been covered by experiments. Current reaction rate uncertainties are difficult to quantify and new measurements are called for. In the absence of more reliable results, we use a hybrid rate consisting of the results from De Smet et al. (2007) and from Rauscher \& Thielemann (2000) below and above $T\approx0.1$ GK, respectively. The rate is listed in the last column of Tab. \ref{tblalrates}. 

\subsection{$^{26}$Al(n,p)$^{26}$Mg ($Q=4.787$ MeV)\label{al26np}}
The $^{26}$Al(n,p)$^{26}$Mg reaction for the transitions to both the ground and first excited $^{26}$Mg state was directly measured by Trautvetter et al. (1986) at a number of neutron energies (corresponding to $T\approx0.36-3.6$ GK). Koehler et al. (1997) measured the $^{26}$Al(n,$p_1$)$^{26}$Mg reaction, i.e., for population of the first excited state in $^{26}$Mg. The latter rates are claimed to be reliable only for $T\leq0.3$ GK. Near the overlap region, $T\approx0.3$ GK, the latter rate exceeds the former rate by a factor of 2 (see discussion in Koehler et al. 1997 for the possible source of the discrepancy). The disagreement cannot be explained by unaccounted transitions, because the Koehler rate {\it exceeds} the Trautvetter rate. In any case, it is shown in both Trautvetter et al. (1986) and in Skelton, Kavanagh \& Sargood (1987) that the $^{26}$Al(n,$p_0$)$^{26}$Mg reaction rate (i.e., for population of the ground state in $^{26}$Mg) is predicted to be much smaller than the (n,p$_1$) rate.

\begin{figure}
\includegraphics[scale=0.5]{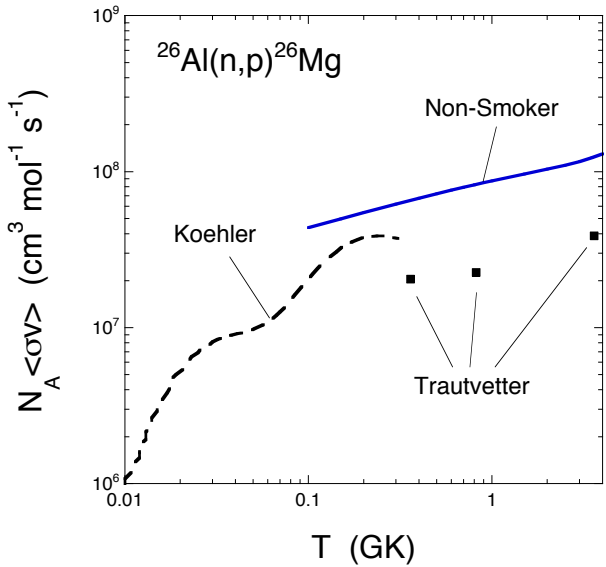}
\caption{Reaction rates for $^{26}$Al(n,p)$^{26}$Mg: (dashed line) Koehler et al. (1997); (blue solid line) Rauscher \& Thielemann (2000); squares (Trautvetter et al. 1986). The first (experimental) rate only takes the transition to the first excited state in $^{26}$Mg into account,
while the third (experimental) rate represents the combined transitions to the ground and first excited states in $^{26}$Mg. The second rate is estimated using the Hauser-Feshbach theoretical model and includes transitions to all possible final states. Note that for this comparison only, the rates represent ``laboratory rates", i.e., they do not account for thermal target excitations. \label{figal26np}}
\end{figure}

The current situation for the reaction rates is displayed in Fig. \ref{figal26np}. The experimental rate for $^{26}$Al(n,p$_1$)$^{26}$Mg from the more recent work of Koehler et al. (1997) is extracted from their Fig. 5 and is shown as a dashed line. The theoretical $^{26}$Al(n,p)$^{26}$Mg rate, based on the Hauser-Feshbach model (Rauscher \& Thielemann 2000), is displayed as a solid blue line. An almost identical theoretical rate has been reported by Goriely, Hilaire \& Koning (2008). Considering the experimental uncertainty of the Koehler rate near their high-temperature cutoff (20\%), the agreement between experimental and theoretical rates near $T\approx0.2$ GK seems reasonable (deviation of 40\%). Therefore, we adopt a hybrid rate consisting of the results from Koehler et al. (1997) and from Rauscher \& Thielemann (2000) below and above $T\approx0.2$ GK, respectively. The rate is listed in column 7 of Tab. \ref{tblalrates}. Note that we prefer the more recent rates from Koehler et al. (1997) and Rauscher \& Thielemann (2000) over the earlier experimental result of Trautvetter et al. (1986), which is displayed as data points in the figure. Our adopted rate exceeds the prediction of Trautvetter et al. (1986) by a factor of $\approx$3 near $T=2.5$ GK. It is currently difficult to estimate rate uncertainties and new measurements are urgently needed. 

\clearpage

\clearpage

\end{document}